\documentclass [10 pt, onecolumn]{article}
\usepackage{amsmath}
\usepackage{txfonts}
\usepackage{amsfonts}
\usepackage{subfigure}
\usepackage{setspace}
\usepackage{booktabs}
\usepackage{amsmath}

\usepackage{mathrsfs}
\usepackage{amsmath}
\usepackage{bbm}
\usepackage{times,amsmath}
\usepackage{multirow}
\usepackage{graphicx}
\usepackage{epsfig}
\usepackage{float}
\usepackage{subfigure}
\makeatletter

\newcommand{\Rmnum}[1]{\expandafter\@slowromancap\romannumeral #1@}
\makeatother

\usepackage{caption}
\usepackage{graphicx, subfig}
\RequirePackage{indentfirst}
\usepackage[justification=centering]{caption}
\date{}

\title{New Computing Model of $GN_eTM$ Turing Machine On Solving Even Goldbach Conjecture}
\author{ Bogang LIN \thanks{College of Computer and Data Science, Fuzhou University, Fuzhou,  China,350108(linbg@fzu.edu.cn). The preliminary research for this work was supported by the National Natural Science Foundation of China (No.60172017) and The Science and Technology Development Project of Fujian Province (No.2007F5071). } }
\begin{document}
\maketitle
\textbf{Abstract } Based on the propositional description of even Goldbach conjecture, in order to verify the truth of even Goldbach conjecture, we will deeply discuss this question and present a new computing model of $G{{N}_{e}}TM$ Turing Machine. This paper proves the infinite existence of even Goldbach's conjecture and obtains the following new results:

1. The criterion of general probability speculation of the existence judgment for even Goldbach conjecture is studied, and which at least have a formula satisfy the deduction result of matching requirements for even Goldbach conjecture in the model $\bmod \overset{\equiv }{\mathop{M}}\,({{N}_{e}})$.

2. In the controller of the $G{{N}_{e}}TM$ model, the algorithm problem of the prime matching rule is given, regardless of whether the computer can be recursively solved, the rule algorithm for designing prime numbers in controllers is computer recursively solvable.

3. The judgment problem that about even Goldbach conjecture whether infinite existence is studied. The new research result has shown that according to the constitution model of the full arranged matrix of given even number ${N}_{e}$, and if only given an even number ${N}_{e}$, it certainly exists the matrix model $Mod\overset{\equiv }{\mathop{X}}\,(p)$ and is proved to be equivalent. Therefore, it proves indirectly that the model $G{{N}_{e}}TM$ does not exist halting problem, and it also indicate that the even Goldbach conjecture is infinity existence.


\section{The posing of the question}

The theory research about even Goldbach conjecture and the computer search has obtained significant achievement in many studies [1-9], but the research of its last certify is little progress. In artificial intelligence age and recent years, we have been known that Simpleminded AlphaGo of powerful intelligence computing wins by a high score to get the better of Legendary players [10], it not only to shock the whole world but also opened a new research idea to us. Is that the computer can solve the problem of the even Goldbach's conjecture? And any given an even number, the computer within finite step whether or not calculable, this is people very concerned one question. If the problem of even Goldbach conjecture is the computer recursion solvable, then it's solution model how to correct description? And how to confirm even Goldbach conjecture whether to exist uniqueness judge result? It still can solve the problem of infinite computation by computer. The question of these key points is discussed in this paper.

Suppose the solution process of the computer for even Goldbach conjecture can be abstracted as new computing model of dream Turing Machine of a simplified deformation, and abbreviated to $G{{N}_{e}}TM$----Goldbach Number Turing Machine,to use it distinguish the results of true (T) or false (F) about the prime matching existence. As shown in Fig 1.

 \begin{figure}
  \centering
  \includegraphics[width=.8\textwidth]{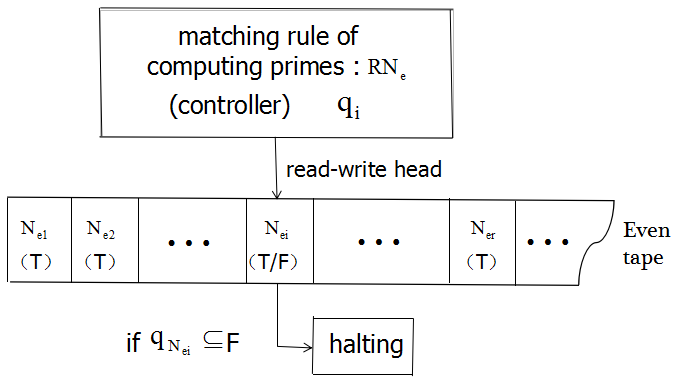} 
  \caption{new computing model of the dream Turing Machine
of a simplified deformation for even Goldbach conjecture
} 
  \label{img} 
\end{figure}
The machine model is consist of some parts of the next contents:

1. A finite controller, i.e., a set of computing control rule $R{{N}_{e}}$ of the prime matching algorithm that to solve even Goldbach conjecture. It is according current read-write head to point at the tape lattice even number ${{N}_{ei}}\;(i=1,2,\cdots )$ and present machine run to be in the state ${{q}_{i}}$ of computation result, in turn, determined read-write head to execute next run operation, and changing the result (T/F) of the status register, even which makes the machine change over to the next new status.

2. An input tape of has infinite tape lattice. The left endpoint is initiation point, right endpoint point to infinite. In each tape lattice just to denote the number of the even set $\{4,6,8,\cdots \}$ that from the number 4 starting. But there is not the number 0 and 2, as well as a special blank mark$\Box$. From the left to the right, in turn, serial number is $1,2,3,\cdots $ in the tape.

3. A read-write head. It can along the left and the right to move in the tape, and to execute  three action: first is to change determine the status of computing result in finite controller; second is to execute normal operation: the function of in order the right, it can read out corresponding even number in present tape lattice. And to execute abnormal operation: the function of in order to the left, and can repeat the operation, several time in succession readout corresponding even number in present tape lattice (use to check). The third is to execute the output result of controller status, and to print out true (T) or false (F) symbol in the corresponding tape lattice.

4. An output status register. It is used to save the computation result status of the present position in the Turing machine, i.e., ${{q}_{i}}\in \{T/F\}$. The numbers of all possible states in the model $G{{N}_{e}}TM$ is limited, it mainly indicates the correct status number of the prime matching and incorrect status number of the prime matching. The latter state is appointed to special halting status.

Here special provision that the halting status of $G{{N}_{e}}TM$ machine model refer to if the prime matching for even Goldbach conjecture is not established, then the computing result makes machine halting (but the process may repeat read the number and check the result). Otherwise, if the machine not halting, then indicate the result of the prime matching is established, and the machine continues to run to infinity.

The model $G{{N}_{e}}TM$ is detailed description as follow:

\textbf{Definition 1.1.} The computer solvable process of even Goldbach conjecture which assumes is called new computing model of dream Turing Machine about a simplified deformation, and abbreviated to $G{{N}_{e}}TM$. TM has a 7-Tuple:
\[\{QG{{N}_{e}},\Sigma {{N}_{e}},R{{N}_{e}},{{q}_{0}},\delta ,\text{T},\text{F}\}\]

$QG{{N}_{e}}$: The infinite set of the non-empty status that about the result both of right and wrong of the prime matching computing $({{p}_{i}}+{{p}_{j}}={{N}_{e}})$ for even Goldbach conjecture. $\forall {{q}_{i}}\in
QG{{N}_{e}},$ ${{q}_{i}}\;(i\ge 1)$ is a status of TM, It is except for original state, or as right status or as wrong status, ${{q}_{i}}\subseteq \{T/F\}$.

${{q}_{0}}$: an original status, ${{q}_{0}}\in QG{{N}_{e}}$. TM is run operation from ${{q}_{0}}$ state start, at the same time, the read-write head point to the leftmost even number in the tape.

$\Sigma {{N}_{e}}$: It is the even number set of finite input: $\{4,6,8,\cdots ,{{N}_{e}}\}\;({{N}_{e}} = 2n,n\ge 2)$, ${{N}_{e}}\in \Sigma {{N}_{e}}$. Only if the even number of given at $\Sigma {{N}_{e}}$ and $\Sigma {{N}_{e}}_{_{{}}}^{\infty }$, it can when TM starting to appear at the input tape. Among other excluding the number 0, 2, and a special blank symbols$\Box$; $\Sigma {{N}_{e}}_{_{{}}}^{\infty }$ is specific reference to input even number set of dream infinite $\{4,6,8,\cdots ,{{N}_{e}},\cdots \}$.

$R{{N}_{e}} $: In the controller of $G{{N}_{e}}TM$, it expresses the rule set of the computing prime matching is the correct or not, and it is the finite matching computation of corresponding each ${{N}_{e}}$ in the independence closed.

$\delta $: The element $\delta :QG{{N}_{e}}\times R{{N}_{e}}\to QG{{N}_{e}}\times R{{N}_{e}}\times \{R,L\}$ is the transition function. Here R and L express read-write head are moved to the right or left.

T(true): It is the matching status that the result appears correct by controller computing, $T\in QG{{N}_{e}}$, ${{q}_{i}}\subseteq T\;(i\ge 1)$. i.e., the machine continues to run status.

F(false): It is the matching states that the result appears incorrect by controller computing, $F\in QG{{N}_{e}}$. If ${{q}_{j}}\notin T\;(j\ge 1)$, If only ${{q}_{j}}\subseteq F\;(j=1)$, then $G{{N}_{e}}TM$ is the termination status, i.e., the machine is halting status.

The run process of the operation status on $G{{N}_{e}}TM$ Turing Machine as shown in Fig 2.
\begin{figure}[htb]
   \centering
   \includegraphics[width=85mm]{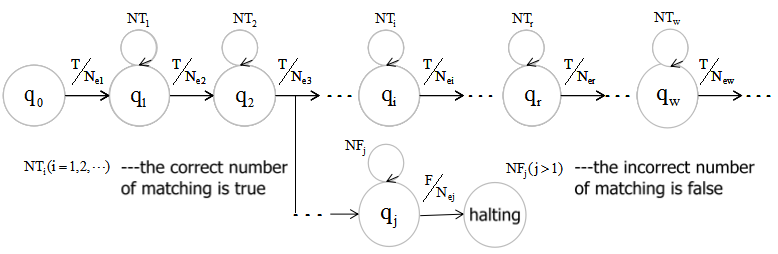}
    \caption{The sketch map of the operation status run on  Turing Machine}
    \label{Machine}
\end{figure}

According to the principle of $G{{N}_{e}}TM$ computing model, we further discussion above model whether exist thus a program (it is used to check the algorithm of the prime matching), it can computing given even $N{}_{e}$ in the input tap, and to can prove that the even Goldbach conjecture is all computer recursion that can be solved. Secondly, it as also can further to verify $\forall {{q}_{i}}\subseteq T\;(i\ge 1)$, $N{{T}_{i}}\;(i=1,2,\cdots )\in T$,the result is all true. And the machine can judge that $\forall {{N}_{e}}$ in the input tape which to satisfy the even Goldbach conjecture is all established. The machine not halting run to infinity. On the contrary, if to verify some status ${{q}_{j}}\subseteq F\;(j>1)$, and by repeat check $N{{F}_{j}}\equiv {{{N}_{e}}}/{2}\;\subseteq F$, the result is all false. Then the machine stops running and shows that it stops, which indicates that even Goldbach conjecture has not been established.

Aimed the questions raised above, we will depth one by one discusses it in next.

\section{Some Preparation Knowledge}
\textbf{Definition 2.1.} Let ${N}_{e}$ be any non-zero even number and ${N}_{e}=2n,n\ge 1$. Assume $x,y\in {{N}_{e}}$ is two positive integers of ${N}_{e}$. If the difference between ${N}_{e}$ and $y$ can be divided by $x$ in the integer field of ${N}_{e}$, then it call ${N}_{e}$ and $y$ are the congruence of module $x$, it is written as:\\
\begin{equation}
 N_e\equiv y(mod \,x) , \;(x,y\in N_e)
\end{equation}
In addition, if the difference between ${N}_{e}$ and $x$ can be divided by $y$, then it call ${N}_{e}$ and $x$ are the congruence of module $y$, and it is written as:
\begin{equation}
 \centering N_e\equiv x(mod \, y) , \;(x,y\in N_e)
\end{equation}

The formula (1) and (2) are collectively called the pairing formula of the even congruence relations.

Because the two formulas are similar, for the convenience of analysis, the following is discussed only in the case of (1).

\textbf{Definition 2.2.} It is called the operating of the congruence independent closure for ${N}_{e}$, it is referred to satisfy the situation ${{N}_{e}}\equiv {{y}^{{}}}(\bmod \,x)$ $(x,y\in {{N}_{e}})$, the operating of all intermediate values and results are the range within ${N}_{e}$. i.e., $[x|({{N}_{e}}-y)]\equiv 1\to {{N}_{e}}|(x+y)\equiv 1$, and to satisfy the smallest positive residual relationship of the congruence by a given module $x$ (or given module $y$).

Obviously, the congruence independent closure in ${N}_{e}$, the definition 2.1 describes the expression relationship of add sum of two even numbers. Another, according to the theory of the congruence, we can easily get the following lemma.

\textbf{Lemma 2.1.} (the congruence expression theorem of implicit even sum )  Assume that ${N}_{e}$ is non-zero even number, ${{N}_{e}}=2n,n\ge 1$, and $x,y=2n-1,n\ge 1\;(x,y\in {{N}_{e}})$, and $x,y$ both are positive odd integer in ${N}_{e}$. In the independent closure of ${N}_{e}$, any given a positive even number ${{N}_{e}}$, it must exists two positive odd integers $x$ and $y$, and to satisfy the as following relation:
\[{N}_{e}\equiv y(mod\, x) \;  (x,y\in N_e)\]

\textbf{Proof.} According to the congruence relation, as long as the independent closure within ${{N}_{e}}$ is given, two positive odd integers $x$ and $y$ will be found to form a congruence relation:

\begin{equation}
\nonumber
x\equiv -y (mod\, N_e) \quad  (x,y\in N_e)
\end{equation}
\begin{equation}
\nonumber
(or: -x\equiv y (mod \,N_e)\quad (x,y\in N_e)
\end{equation}
Are established and satisfy $({{N}_{e}}|(|x+y|))\equiv 1$.

Now think about it the other way around, by definition 2.1, if only for the element $x$, swap ${N} _ {e}$ with $x$, and y is fixed. Because $x$ and $y$ are both positive odd numbers in the element $x \not = y $ or $x = y $. According to the congruence relation, the following congruence expressions with even numbers can be summed, i.e.,
\[N_e\equiv y(mod\, x) \quad  (x,y\in N_e)\]
It is established, and satisfied the $[x|(({{N}_{e}}-y)=x)]\equiv 1 $. For have $y={{N}_{e}}-x$, above result can be also expressed as:
\[N_e\equiv N_e-x(mod\, x)\]
And to satisfy the formula as: $[x|({{N}_{e}}-({{N}_{e}}-x))]\equiv 1$.

In fact, the lemma of this implicit expression even sum, which actually describes the establishment result of ${N}_{e}\equiv y+x$. The lemma shows that an even Goldbach Guess is the relationship between adding the sum of two odd numbers. Lemma 2.1 can also describe equivalently the existence relation model of even add sum within ${N}_{e}$ independent closure.

\textbf{Definition 2.3.} `` The odd complete congruence expression group of determinate module $x$ " means that within the congruence independent closure of ${N}_{e}$, even number ${N}_{e}$ is the relationship formulas of the congruence sum of consisted of all enable it established full permutation for two odd positive integers element $x$ and $y$, i.e.,
\begin{equation}
\nonumber
\left\{\begin{array}{l}{{N}_{{e}} \equiv {y}_{{N}_{e}-1} \bmod {x}_{1}} \\
{N}_{{e}} \equiv {y}_{{N}_{e}-3} \bmod {{x}_{3}} \\
{\vdots} \\
{{N}_{{e}} \equiv {y}_{3} \bmod {x}_{{N}_{e}-3}} \\ {{N}_{{e}} \equiv {y}_{1} \bmod {x}_{{N}_{e}-1}}\end{array}\right.
 \end{equation}

And it's written as:
  \begin{equation}
 \nonumber
           {{N}_{e}}{{(o)}^{{\;}}}{{\underline{{\bar{\equiv }}}}_{{}}}{{Y}_{o}}^{{\;}}(\bmod \,{{X}_{o}}) \quad (short note:Mod\,\overset{\equiv }{\mathop{X}}\,(o))\\
  \end{equation}

\textbf{Lemma 2.2.} Independent closure in ${{N}_{e}}$, ``The odd complete congruence expression group of determinate module $x$ " contains full permutation congruence expression of the odd integrity of ${{N}_{e}}$/2 pairs, and it satisfies the basic characters: uniqueness, closure, symmetry,  constructive and expansible.

\textbf{Proof.} Selecting ${{k}^{{}}}(1\le k\le {{N}_{e}}-1)$ as an enumerate factor within the congruence independent closure of ${{N}_{e}}$, after to act on $x$ and $y$ by $k$ separately, always existent respective formula ${{N}_{e}}\;{{\equiv }_{{\;}}}{{y}^{'}}(\bmod\, {{x}^{'}})$, and makes it is established. And the structure of full permutation congruence expression for the minimum positive remaining of all odd integer, which as showed in the definition 2.3. i.e.,
\begin{equation}
 \nonumber
                 {{N}_{e}}{{\;\overline{\underline{\equiv }}}^{{\;}}}{{Y}_{{{k}^{'}}}}(\bmod\, {{X}_{k}})={{[{{N}_{e}}\equiv {{y}_{{{N}_{e}}-k}}{{\bmod\, }_{k}}]}_{(1\le k\le {{N}_{e}}-1)}}
\end{equation}

Here ${{k}^{'}}$ and $k$ are mutual duality relation. Obviously, through the determinate the congruence expressions shape of the integrity full permutation within the independent closure of ${{N}_{e}}$, when $x,y=2n-1,(n\ge 1)$, for $1\le k\le {{N}_{e}}-1$, as long as the enumerate of $k$ is not lack  the number of items. According to the definition 2.1 and lemma 2.1, in the complete congruence expression group for all determinate module $x$, which must exist full permutation form of integrity congruence expressions of ${{N}_{e}}/2$ pairs. And it satisfies as following the trait.

\textbf{Uniqueness.} Given any even number ${{N}_{e}}$, and enumerating each number at $1\le k\le {{N}_{e}}-1$, the module $x$ from $1\to {{N}_{e}}-1$ enumerating, the element $y$ is corresponding $x$ by ${{N}_{e}}-1\to 1$ with it paired, and it satisfies the relation permutation of ${{[{{N}_{e}}\equiv {{y}_{{{N}_{e}}-k}}\bmod  {{x}_{k}}]}_{(1\le k\le {{N}_{e}}-1)}}$. As long as ${{N}_{e}}$ is given, and the form that full permutation not lack the number of items of the odd integrity congruence expression group is also assured, then the uniqueness is permanently true.

\textbf{Closure.} Examining every congruence expression ${{N}_{e}}^{{\;}}{{\underline{{\bar{\equiv }}}}_{{\;}}}{{Y}_{{{k}^{'}}}}^{{}}(\bmod \, {{X}_{k}})$,
${[{N}_{e}\equiv {y}_{{{N}_{e}}-k}}$
mod ${{x} _{k}}]$, $(1\le k\le {{N}_{e}}-1)$ are all established, and it satisfy to $[{{x}_{k}}|({{N}_{e}}-{{y}_{{{N}_{e}}-k}})]\equiv 1$. According to the definition 2.2. All the congruence expressions are operating in the independent closure in ${{N}_{e}}$. Then, by lemma 2.1 can be known that its closure is really.

\textbf{Symmetry.} From the shape of the full permutation and the character of the uniqueness of the integrity congruence expression, it's easy found the existence of the symmetry relationship in $Mod\,\overset{\equiv }{\mathop{X}}\,(o)$. Namely, ${{N}_{e}}\;\underline{\overline{\equiv }}\;{{Y}_{{{k}^{'}}}}(\bmod \,{{X}_{k}})=$ ${{[{{N}_{e}}\equiv {{y}_{k}}\bmod\, {{x}_{{{N}_{e}}-k}}]}_{(1\le k\le {{N}_{e}}-1)}}$ is true too. If only $k$ just selecting to a half of the range can satisfy the symmetrical relationship.

\textbf{Constructability.} When ${{N}_{e}}$ is given, then ${{N}_{e}}^{{\;}}{{\underline{{\bar{\equiv }}}}_{{\;}}}{{Y}_{{{k}^{'}}}}^{{}}( \bmod\, {{X}_{k}})$ is constructed by the effect of the enumerate factor $k$. As long as enumerate $k$ makes all congruence expression group is established in the range of $1\le k\le {{N}_{e}}-1$, theoretically, in $\forall {{N}_{e}}$, we enumerating every ${{N}_{e}}$, it satisfies the construction of the full permutation of the odd integrity congruence expression group. It obviously establishes.

\textbf{Expansible.} If only the element $x,y\to \infty $,${{N}_{e}}\to \infty $, then $Mod\,{{}^{{}}}\overset{\equiv }{\mathop{X}}\,(o)\to \infty $. The result must be existing expansible.

\textbf{Deduction 2.1.} Assume that $\phi (Mod\,\overset{\equiv }{\mathop{X}}\,(o))$ is expressed as the numbers of full arrangement pairs (allowing equivalent repeatable) that the congruence relation consist of all odd number ($x,y$) to exist all determined modulo x in $Mod\,\overset{\equiv }{\mathop{X}}\,(o)$, then there is $\phi (Mod\,\overset{\equiv }{\mathop{X}}\,(o))={{N}_{e}}/2$.

\textbf{Proof.} According to the definition 2.3 and the lemma 2.2, the deduction established is can be proved.

\textbf{Definition 2.4.} ``The congruence expression group of the prime of determinate module $x$" means that in the independent closure of ${{N}_{e}}$, the element $x,y$ are all consist of the prime $(p{{'}_{1}},p{{'}_{2}},p{{'}_{3}},\ldots ,p{{'}_{s-2}},p{{'}_{s-1}},p{{'}_{s}})$ $(1\le s<r)$, it is the set of the congruence add sum expression of a set of the permutation relation in determined module $x$. It is written as follow:
\begin{equation}
 \nonumber
         {{N}_{e}}{{(p)}^{{\;}}}{{\underline{{\bar{\equiv }}}}_{{\;}}}Y{{_{p}^{{}}}^{{}}}(\bmod\, X_{p}^{{}})\quad  (short note: Mod\overset{\equiv }{\mathop{X}}\,(p))
 \end{equation}

In here, $Mod\,\overset{\equiv }{\mathop{X}}\,(p)$ $\in Mod\,\overset{\equiv }{\mathop{X}}\,(o)$. Because of $Mod\,\overset{\equiv }{\mathop{X}}\,(p)$ in this definition, among the element $x,y$ are all consist of the primes ${{p}_{i}}$ $(1\le i\le r)$, but it must satisfy the result establish of the formula: ${{X}_{p}}+{{Y}_{p}}={{N}_{e}}(p)$, it's not all prime in ${{N}_{e}}$ can be constituted the matching and to can satisfy this relation of even add sum.

\textbf{Definition 2.5.} ``The prime congruence formulas of regular determinate module $x$" are expressed as the set of the congruence formulas of the prime universal arrangement in modulo $x$, which element can construct a group of non-repeating relation result in $Mod\,\overset{\equiv }{\mathop{X}}\,(p)$. It is written as the formula as following:
\begin{equation}
 \nonumber
N_{e}^{+}{{(p)}^{{\;}}}{{\underline{{\bar{\equiv }}}}_{{\;}}}Y{{_{p}^{+}}^{{}}}(\bmod \,X_{p}^{+}), (short note: Mod\,\overset{\equiv }{\mathop{{{X}^{+}}}}\,(p))
\end{equation}

Here, $Mod\,\overset{\equiv }{\mathop{{{X}^{+}}}}\,(p)$ is also a kind of set that corresponding positive congruence residue system in the independent closed module $x$, among the element $x,y$ are all the prime and one to one non-repeating composing the congruence relation result. Therefore, $Mod\,\overset{\equiv }{\mathop{{{X}^{+}}}}\,(p)$ without the symmetry, but it still satisfies uniqueness, closed and constructible, as well as Expansible. So the relation as follow:
\[Mod\,\overset{\equiv }{\mathop{{{X}^{+}}}}\,(p)\in Mod\,\overset{\equiv }{\mathop{X}}\,(p)\in Mod\,\overset{\equiv }{\mathop{X}}\,(o)\]
\[Mod\,\overset{\equiv }{\mathop{{{X}^{+}}}}\,(p)\in Mod\,\overset{\equiv }{\mathop{{{X}^{+}}}}\,(o)\in Mod\,\overset{\equiv }{\mathop{X}}\,(o)\]

\textbf{Lemma 2.3 (Chinese remainder theorem).} Assume that ${{m}_{1}},{{m}_{2}},\cdots ,{{m}_{r}}$ are r positive prime of the coprime numbers, ${{b}_{1}},{{b}_{2}},\cdots ,{{b}_{r}}$ are any r positive integers, then have r congruence equations:
\begin{equation}
 \nonumber
                       S\equiv b_j( mod\, m_i)\quad(1\le i,j\le r)
 \end{equation}

Corresponding the module $m={{m}_{1}}{{m}_{2}}\cdots {{m}_{r}}$, it has the result of unique one solution. The solution is:
      \[ S\equiv M_{1}^{'}{{M}_{1}}{{b}_{1}}+M_{2}^{'}{{M}_{2}}{{b}_{2}}+...+M_{r}^{'}{{M}_{r}}{{b}_{r}}(\bmod \, m)\equiv \sum\limits_{i,j=1}^{r}{M_{i}^{'}{{M}_{i}}{{b}_{j}}}(\bmod \, m)\]

Where $m=m{}_{1}m{}_{2}...m{}_{r}$,${{M}_{i}}={}^{m}/{}_{{{m}_{i}}}$,$M_{i}^{'}{{M}_{i}}\equiv 1\left( \bmod\, {{m}_{i}} \right)$,$M_{i}^{'}=M_{i}^{-1}\left( \bmod\, {{m}_{i}} \right)$, $(1\le i\le r)$.(Proof omitted)

Lemma 2.3 describes the summation problem of r unknown congruent equations, where ${{b}_{j}}$ and ${{m}_{i}}$ is known, and $S$ is unknown. If now let $S$ and ${{m}_{i}}$ is known, ${{b}_{j}}$ is unknown, then the equation becomes to solve that the problem of ${{b}_{j}}\;(1\le j\le r)$ whether or not can satisfy the solution of the equation. So a new definition is given as follow:

\textbf{Definition 2.6.} In the matching formula of the congruence relations of even add sum, if ${{N}_{e}}=2n, n>2$ is any non-zero even number, let ${{m}_{i}}={{p}_{i}}$ $(1\le i\le r)$, $2<{{p}_{i}}\le {{N}_{e}}-1$, it is any a prime within all odd prime numbers of less than ${{N}_{e}}$, and ${{b}_{j}}\;(1\le j\le r)$ is the positive integers of even congruence matching relations in closed ${{N}_{e}}$. If in closed even ${{N}_{e}}$, selecting every ${{m}_{i}}$ is all have unique ${{b}_{j}}$ corresponds to it, and have all formulas of congruence relation of even matching as follow:
\begin{equation}
 \nonumber
                       \bmod \overset{\equiv }{\mathop{M}}\,({{N}_{e}})=\left\{ \begin{matrix}
   {{N}_{e}}\equiv {{b}_{1}}(\bmod\, {{p}_{1}})  \\
   {{N}_{e}}\equiv {{b}_{2}}(\bmod\, {{p}_{2}})  \\
   ...  \\
   {{N}_{e}}\equiv {{b}_{r}}(\bmod\, {{p}_{r}})  \\
\end{matrix} \right.\\
\end{equation}

(Or it's written as: $\bmod\, \overset{\equiv }{\mathop{M}}\,({{N}_{e}})$ = [${{N}_{e}}$  $\equiv {{b}_{j}}(\bmod\, {{p}_{i}})(1\le i,j\le r)],({{b}_{j}},{{p}_{i}}\in {{N}_{e}})$) Then it is called extended remainder theorem equations, and short note: $\bmod \overset{\equiv }{\mathop{M}}\,({{N}_{e}})$.

The difference between extended remainder theorem equations  $\bmod\, \overset{\equiv }{\mathop{M}}\,({{N}_{e}})$ with lemma 2.3 are that, ${{N}_{e}}$ and ${{m}_{i}}={{p}_{i}}$ are known, ${{b}_{j}}$ is unknown, where ${{b}_{j}}={{N}_{e}}-{{p}_{i}}\;(1\le i,j\le r)$, and each formula can be obtained result through specific solution and computing.

In fact, the definition 2.6 is another form of the definition 2.3, that is the module is the odd number becomes the module is the prime, both relation are $\bmod \, \overset{\equiv }{\mathop{M}}\,({{N}_{e}})$ $\in $ $Mod\,\overset{\equiv }{\mathop{X}}\,(o)$. In the definition 2.6, the element $x\to p$,$y\to b$, x is to consist of all prime ${{p}_{i}}\;(1\le i\le r)$ within ${{N}_{e}}$, another y consist of non-deterministic corresponding quasi-prime $p_{i}^{}\;(1\le i\le r)$ of equal number, $p_{i}^{'}$ is the prime or the odd. In the independent closed ${{N}_{e}}$, practical two prime add result can make the relation of the even sum is established combination pair number, which at most for only $s$ $(1\le s<r)$ pairs. Whether or not each event in ${{N}_{e}}$ at least all has one pair establishes, now to need we further continue verifying it.

And due to the relation of the definition 2.6, $\bmod \, \overset{\equiv }{\mathop{M}}\,({{N}_{e}})\in Mod\,\overset{\equiv }{\mathop{X}}\,(o)$, therefore have:
              \[Mod\overset{\equiv }{\mathop{X}}\,(p)\in \bmod\, \overset{\equiv }{\mathop{M}}\,({{N}_{e}})\in Mod\,\overset{\equiv }{\mathop{X}}\,(o)\]

\section{ The judgment of general probability speculation for even Goldbach conjecture\protect\footnote{The related content of the section 3 and section 4 are to reference the following paper: "Bogang Lin, New Model Depiction of Computer Solvable Proof on Even Goldbach Conjecture, in Proceedings of the China Computer'09 Conference, Tian Jin,China,393-409". }} 

First convention: It can be expressed as even number ${{N}_{e}}$ of two prime add sum called Goldbach number, which is written as $G({{N}_{e}})$. For the convenience of discussion, some concepts are proposed as follows:

\textbf{Definition 3.1.} Assume ${{F}_{1}}=\{{{O}_{s}},{{O}_{is}},{{S}_{o}}\}$ be a class of the patterns set for two odd number sequence corresponding each element arrangement composed of computing add sun in triples element $\{{{O}_{s}},{{O}_{is}},{{S}_{o}}\}$, where ${{O}_{s}}=\{1,3,5,\cdots ,2n-5,2n-3,2n-1\}$ is a permutation set of odd number sequence (from small to large array) within any given even number ${{N}_{e}}=2n\;(n>1)$, ${{O}_{is}}=\{2n-1,2n-3,2n-5,\cdots ,$$5,3,1\}$ is a permutation set of reverse odd number sequence (from large to small array) within any given even number ${{N}_{e}}=2n\;(n>1)$, and ${{S}_{o}}=\{(1+(2n-1))=2n,(3+(2n-3))=2n,\cdots ,$ $((2n-3)+3)=2n,$ $((2n-1)+1)=2n\}$ is the set of even sum that corresponding to each group two element add sum operation.

\textbf{Definition 3.2.} Assume ${{F}_{2}}=\{{{P}_{s}},P_{is}^{'},{{G}^{'}}(N{}_{e})\}$ be a class of the set for the prime and quasi-prime (where have prime or odd number) two sequence composed of corresponding element permutation and each group two element add sun in triples element $\{{{P}_{s}},P_{is}^{'},{{G}^{'}}({{N}_{e}})\}$, where ${{P}_{s}}=\{{{p}_{1}},{{p}_{2}},{{p}_{3}},...,{{p}_{r-2}},$${{p}_{r-1}},{{p}_{r}}\}$ is a permutation set of all prime sequence (from Small number to large number array) within any given even number ${{N}_{e}}=2n\;(n>1)$, $P_{is}^{'}=\{p_{r}^{'},p_{r-1}^{'},p_{r-2}^{'},...,p_{3}^{'},p_{2}^{'},p_{1}^{'}\}$ is a permutation set of reverse quasi-prime sequence (from large number to small number array) within any given even number ${{N}_{e}}=2n\;(n>1)$, and ${{G}^{'}}({{N}_{e}})=\{({{p}_{1}}+p_{r}^{'})={{N}_{e}},\;({{p}_{2}}+p_{r-1}^{'})={{N}_{e}},\cdots ,$ $({{p}_{r-1}}+p_{2}^{'})={{N}_{e}},\;({{p}_{r}}+p_{1}^{'})={{N}_{e}}\}$ is an array set of corresponding to each group of prime and quasi-prime both two element add sum, which is also called quasi- Goldbach number.

\textbf{Definition 3.3.} Let $\varphi ({{S}_{o}})$ be the numbers of the combination pair of two element add sum  for any given even number in ${{F}_{1}}$, $\phi ({{G}^{'}}({{N}_{e}}))$ be the numbers of the combination pair of two element sum relation for any given even number in ${{F}_{2}}$, and $\varphi (G({{N}_{e}}))$ be the numbers that satisfy Goldbach number combination pair of two element add sum in ${{F}_{2}}$.

Further analysis shows that ${{F}_{1}}\ne {{F}_{2}},\;{{F}_{2}}\in {{F}_{1}},\;{{N}_{e}}=2n,\; G({{N}_{e}})\;\in {{G}^{'}}({{N}_{e}}),\;\varphi ({{S}_{o}})\equiv n,\;\varphi (G({{N}_{e}}))\in \varphi ({{G}^{'}}({{N}_{e}})),\;\varphi ({{G}^{'}}({{N}_{e}}))=r,\;\varphi (G({{N}_{e}}))$ are unknown. If $\varphi (G({{N}_{e}}))=0$ (only if can find out an example), then the even Goldbach conjecture is not established. Conversely, if $\varphi (G({{N}_{e}}))\ne 0$, and can ensure $\forall{{N}_{e}},$ $\varphi (G({{N}_{e}}))=1$, then the even Goldbach conjecture is established. This paper will further discuss $\varphi (G({{N}_{e}}))=?$

According to the definition 2.6, a new description model $\bmod\, \overset{\equiv }{\mathop{M}}\,({{N}_{e}})$ of equivalent existence for even Goldbach conjecture has been established. By calculating the probability of ${{q}_{j}}$  and matching to determine the probabilities, we verified the general conjecture criterion of even number conjecture.

\textbf{Theorem 3.1.} Even Goldbach Guess can be implicitly described by $\bmod \,\overset{\equiv }{\mathop{M}}\,({{N}_{e}})$ model.

\textbf{Proof.} By the definition 2.1 and the lemma 2.1, as well as the definitions 2.2 and 3.2, we have known that $\bmod \,\overset{\equiv }{\mathop{M}}\,({{N}_{e}})$ is an extended remainder theorem equation. If ${{N}_{e}}$ is given, in the independent congruence closure of ${{N}_{e}}$, the module ${{m}_{i}}$ $(1\le i\le r)$ is given, too. According the relationship of the even sum, it's easy to obtain the corresponding ${{b}_{j}}$, for ${{N}_{e}}-{{m}_{i}}={{b}_{j}}$. Now let's assume ${{m}_{i}}={{p}_{i}}$ $(1\le i\le r)$ be any element of the prime set within ${{N}_{e}}$, and let ${{m}_{i}}={{p}_{i}}{{,}^{{}}}{{m}_{j}}={{p}_{j}}$, $({{p}_{i}},{{p}_{j}})=1$ be relatively prime. Furthermore, let ${{b}_{j}}={{q}_{j}}\;(1\le j\le r)$ be any odd of among odd integers set that  makes two element add sum relation can established of corresponding to module ${{m}_{i}}={{p}_{i}}$ in ${{N}_{e}}$, ${{q}_{j}}$ be odd (${{q}_{j}}=2n-1$, $n\ge 1$) either prime (${{q}_{j}}={{p}_{i}}$, ${{q}_{j}}\ne {{p}_{i}}$). The result is:
\[\left\{ {\begin{array}{*{20}{c}}
{{N_e} \equiv {q_1}^{}(\bmod\, {p_1})}\\
{{N_e} \equiv {q_2}^{}(\bmod\, {p_2})}\\
{...}\\
{{N_e} \equiv {q_r}^{}(\bmod\, {p_r})}
\end{array}} \right.\]
And it can also be written as: ${{N}_{e}}\equiv {{q}_{j}}(\bmod \,{{p}_{i}})\;(1\le i,j\le r), ({{q}_{j}},{{p}_{i}}\in {{N}_{e}})$

We use mathematical induction to prove that the result is established. When ${{N}_{e}}$= 6,  then $6\equiv {{3}^{{}}}(\bmod \,3)$ is established. When ${{N}_{e}}=m=$ $2n+2\ge 6\;(n\ge 2)$, then $m\equiv {{q}_{mj}}(\bmod \,{{p}_{mi}})$ $(1\le i,j\le r)$, $({{q}_{mj}},{{p}_{m}}_{i}\in m\;=\;{{N}_{e}})$ are established. Furthermore, when ${{N}_{e}}=m+2$, $(m=2n+2\ge 6,n\ge 2)$, then  $m+2\equiv {{q}_{(m+2)j}}(\bmod \, p{}_{(m+2)i})$ $(1\le i,j\le r)$,  $({{q}_{(m+2)j}},\;{{p}_{{{(m+2)}_{{}}}i}}$ $\in m+2={{N}_{e}})$ is established.

Obviously, this structure model gives a new implicit description model of even Goldbach Guess, and each ${{N}_{e}}$ is all may be used the structure to describe it. If only ${{N}_{e}}$ is given, it is can be any enumeration. Once ${{p}_{i}}\;(1\le i\le r)$ is determined, ${{q}_{j}}\;(1\le j\le r)$ by solution result can also get determined corresponding element. Because ${{q}_{j}}={{N}_{e}}-{{p}_{i}}\;(1\le i,j\le r)$ can be determined to be an odd or a prime number, so we use the model $\bmod \,\overset{\equiv }{\mathop{M}}\,({{N}_{e}})$ to describe the Even Goldbach Guess is also unique.

Now for example are described below, suppose ${{N}_{e}}=34$, the prime set $P={\{3,5,7,11,}$
${13,17,19,23,29,31\}}$ within ${{N}_{e}}$, then have odd integers set $Q={\{31,29,27,23,}$
${21,17,15,11,5,3\}}$ of corresponding in ${{N}_{e}}$. The form of $\bmod \, \overset{\equiv }{\mathop{M}}\,({{N}_{e}})$ is given as following:
\[\bmod \mathop M\limits^ \equiv  (34) = \left\{ {\begin{array}{*{20}{c}}
{34 \equiv 31\bmod {3_{}}}\\
{34 \equiv 29\bmod {5_{}}}\\
{34 \equiv 27\bmod {7_{}}}\\
{34 \equiv 23\bmod 11}\\
{34 \equiv 21\bmod 13}\\
{34 \equiv 17\bmod 17}\\
{34 \equiv 15\bmod 19}\\
{34 \equiv 11\bmod 23}\\
{34 \equiv {5_{}}\bmod 29}\\
{34 \equiv {3_{}}\bmod 31}
\end{array}} \right.\]

Now we need to solve a question , if $\forall {{N}_{e}}$, each ${{N}_{e}}$ corresponding the element ${{q}_{j}}$ within Q ($({{q}_{j}}\in Q(1\le j\le k,Q\in {{N}_{e}}))$ sure exist a prime, and satisfy the relation of ${{q}_{j}}={{N}_{e}}-{{p}_{i}}\;(1\le i,j\le r)$, then the result definitely at least have a formula satisfy the requirement of even Goldbach conjecture in $\bmod\, \overset{\equiv }{\mathop{M}}\,({{N}_{e}})$. Thus it's easy to verify the result of even Goldbach conjecture.

According to the definition 3.2, the theorem 3.1, and the analysis of the construction trait for the equation group $\bmod\, \overset{\equiv }{\mathop{M}}\,({{N}_{e}})$ of Expanded Chinese Remainder Theorem(ECRT) in the definition 2.6, we have been found that, for even Goldbach conjecture can be expressed using the model $\bmod\, \overset{\equiv }{\mathop{M}}\,({{N}_{e}})$. When given any even number ${{N}_{e}}=2n,n>2$, within the independence closed range of every ${{N}_{e}}$, the odd prime set $P=\{{{p}_{i}}\}\;(1\le i\le r)$ is defined, too. For all odd prime not greater than ${{N}_{e}}$ as follow:
                   \[3={{p}_{1}}<{{p}_{2}}<{{p}_{3}}<\cdots <{{p}_{r}}<{{N}_{e}}\]

With it correspondent odd integer set $Q={{q}_{j}}\;(1\le j\le r=k)$ is also can be determined uniquely, because of
 ${{q}_{j}}={{N}_{e}}-{{p}_{i}}\;(1\le i,j\le r)$. Now the key question is whether we can prove that at least one of the elements  ${{q}_{j}}$ in $Q$ is a prime number? and satisfy the relation result of ${{q}_{j}} = {{N}_{e}}-{{p}_{i}}\;(1\le i,j\le r)$. That is to say, If its existence can prove to be a definite truth, then the even conjecture question can be solved. On the point, first, we further analysis  the model described in the definition 3.2, 3.3 and the theorem 3.1, and for it use simplifying handle, main consider key structure in $\bmod\, \overset{\equiv }{\mathop{M}}\,({{N}_{e}})$ as following:
\[{G_{ij}}'({N_e}) = {\left\{ {\begin{array}{*{20}{c}}
{{p_i}}\\
{{q_j}}\\
{{N_e}}\\
\end{array}} \right\}_{}}(1 \le i,j \le r = k)\]

Therein, if ${{p}_{i}}$ and ${{q}_{j}}$ is a prime number and satisfy matching requiring of the even established relation of two prime add rum, then ${{N}_{e}}={{p}_{i}}+{{q}_{j}}$ is Goldbach number $G({{N}_{e}})$. We special agreement, if ${{N}_{e}}=2n,n>2$ is any given even, ${{p}_{i}}\in {{N}_{e}}$ is determined prime, ${{q}_{j}} $ is only correspond the matching prime of even established relation of two element add sum, i.e., ${{q}_{j}}={{N}_{e}}-{{p}_{i}}\; (1\le i,j\le r)$, then ${{G}_{ij}}^{'}({{N}_{e}})$ is can called satisfy with pairing requirements of Goldbach number $G({{N}_{e}})$. Otherwise, ${{G}_{ij}}^{'}({{N}_{e}})$ is called does not satisfy the matching requirements of Goldbach number $G({{N}_{e}})$.

We have obtained new results by the proof as following:

\textbf{Theorem 3.2.}  In the model $\bmod\, \overset{\equiv }{\mathop{M}}\,({{N}_{e}})$, any given an even number ${{N}_{e}}$, once the element of within the prime set P is determined, by randomly selecting the element ${{q}_{j}}$ within Q,if only at $r\approx 0.8\text{33}\sqrt{{{N}_{e}}}$ computing the number value, then the element ${{q}_{j}}$  in the probability of 50\% get one element is the prime, and ${{G}_{ij}}^{'}({{N}_{e}})$ satisfy the matching requirements of Goldbach's number $G({{N}_{e}})$.

\textbf{Proof.} Assume ${{N}_{e}}$ be any big enough even number, ${{N}_{e}}=2n,n>2$. In addition assume $N$ as a big enough integer, and ${{N}_{e}}\subseteq N,N=\{1,2,3,\cdots ,{{N}_{e}}-1,{{N}_{e}}\}$. In the model $\bmod \, \overset{\equiv }{\mathop{M}}\,({{N}_{e}})$, because of the set $P$ = {${{p}_{1}},{{p}_{2}},\cdots ,{{p}_{r}}$} of odd prime in the independent closure of $N$ is may be determined, and the set $Q$ = {${{q}_{1}},{{q}_{2}},...,{{q}_{r}}$} of corresponding odd integers can also be determined uniquely, for $Q\in N,Q = N-P$. Any an element ${{q}_{j}}$ must with another element ${{p}_{i}}$ forming one-to-one mapping to a matching relationship. For ${{p}_{i}}$ be the prime, ${{q}_{j}}$ is either odd or prime. Now it can be considered in this wise, because of all the primes in P are determined, it means that every prime within the r box in $P$ has been selected. What is corresponding the element ${{q}_{j}}$ in $Q$? It is either an odd number or a prime. That is to say, how many is the number of the prime that it is exactly a prime and satisfying the requirements of $G({{N}_{e}})$? We convert directly the problem into a solution ${{G}_{ij}}^{'}({{N}_{e}})$  whether or not satisfy the matching requirements of Goldbach number $G({{N}_{e}})$ for analysis.

Now might as well assume that $P_{G}^{j}(1\le j\le r)$ choose a different scheme from each other that randomly select anyone matching element within ${{q}_{j}}\;(1\le j\le r)$ in N, and the result makes ${{G}_{ij}}^{'}({{N}_{e}})$ all not satisfy the matching requirements of Goldbach number $G({{N}_{e}})$[11]. Here's appointed: those have been selected element is no longer a part of option again, too.

We firstly consider ${{q}_{1}}$ can arbitrarily select an element in $N$, ${{G}_{ij}}^{'}({{N}_{e}})$ not satisfy the probability of the requirement of $G({{N}_{e}})$ is ${\scriptstyle{}^{1}/{}_{N}}$. The second let ${{q}_{2}}$ choose another an element in $N$, and ${{q}_{2}}$ $\ne $ ${{q}_{1}}$, and that when selecting ${{q}_{2}}$, ${{G}_{ij}}^{'}({{N}_{e}})$ not satisfy the probability of the requirement of $G({{N}_{e}})$ is $(1-{\scriptstyle{}^{1}/{}_{N}})$. Third let ${{q}_{\text{3}}}$ choose the remaining elements in $N$, and ${{q}_{\text{3}}}$ $\ne $ (${{q}_{1}}$,${{q}_{2}}$), and that when selecting ${{q}_{3}}$, ${{G}_{ij}}^{'}({{N}_{e}})$ not satisfy the probability of the requirement of $G({{N}_{e}})$ is $(1- {\scriptstyle{}^{2}/{}_{N}})$,..., and so on. At last, let ${{q}_{\text{r}}}$ choose the remaining elements in $N$, and ${{q}_{\text{r}}}$ $\ne $ ($ {{q}_{1}}$,${{q}_{2}}$,${{q}_{\text{3}}}$,...,${{q}_{\text{r-1}}}$), and that when selecting ${{q}_{\text{r}}}$, and ${{G}_{ij}}^{'}({{N}_{e}})$ not satisfy the probability of the requirement of $G({{N}_{e}})$ is $(1- {\scriptstyle{}^{r-1}/{}_{N}})$.

Because ${{q}_{j}}={{N}_{e}}-{{p}_{i}}\;(1\le i,j\le r)$  must to determine one ${{q}_{j}}$ element, so ${{q}_{j}}$ within $r$ by random selecting different the prime, either of ${{q}_{j}}$  all not appears it is the prime, ${{G}_{ij}}^{'}({{N}_{e}})$ not satisfy the matching requirement of Goldbach number $G({{N}_{e}})$; either of ${{q}_{j}}$ even if to appear it is the prime, but ${{G}_{ij}}^{'}({{N}_{e}})$ still not satisfy the matching requirement of the Goldbach number $G({{N}_{e}})$. Obviously, the probability of thus not satisfied condition that described as follow.
\[(1 - \frac{1}{N})(1 - \frac{2}{N}){(1 - \frac{{\rm{3}}}{N})^{}}{...^{}}(1 - \frac{{r - 1}}{N}) \approx \prod\limits_{i = 1}^{r - 1} {(1 - \frac{i}{N}} )\]

Here, we can let $x={}^{i}/{}_{N}$. When $N$ is a large integer, and $x$ is a small real number, according to the following relation of the formula:
\[{e^{ - x}} = 1 - x + \frac{{{x^2}}}{{2!}} - \frac{{{x^{\rm{3}}}}}{{{\rm{3!}}}} +  \cdots \]

Then have $1-x\approx {{e}^{-x}}$. All things considered, the selecting results if $r$ primes does not appear in the $r$ selecting scheme, or ${{q}_{j}}$ even if to appear be prime, but ${{G}_{ij}}^{'}({{N}_{e}})$ still not satisfy the probability of the matching requirement of Goldbach number $G({{N}_{e}})$, then have the following result:
\[\prod\limits_{i = 1}^{r - 1} {(1 - \frac{i}{N}} ){ \approx ^{}}\prod\limits_{i = 1}^{r - 1} {{e^{ - \frac{i}{N}}}}  = {e^{ - \frac{{r(r - 1)}}{{2N}}}}\]

Now, we conversely consider problem can find that, in r selecting scheme, ${{q}_{j}}$ may be selected as the prime more than once in $r$ options, but at least have one prime makes ${{G}_{ij}}^{'}({{N}_{e}})$ to satisfy the requirements of Goldbach number $G({{N}_{e}})$. Or selecting right only one element is a prime, but it can match ${{G}_{ij}}^{'}({{N}_{e}})$ which satisfies  the requirement of the Goldbach number $G({{N}_{e}})$, the probability of the request as follow.
\[{P_G} =1-\prod\limits_{i = 1}^{r - 1} {(1 - \frac{i}{N}} ){ \approx 1- ^{}}\prod\limits_{i = 1}^{r - 1} {{e^{ - \frac{i}{N}}}}  =1- {e^{ - \frac{{r(r - 1)}}{{2N}}}}\]
Let's assume:
\[{P_G} = \theta  = 1 - {e^{ - \frac{{r(r - 1)}}{{2N}}}}\]

After by proper arrangement, there is ${{r}^{2}}-r=N\ln \frac{1}{{{(1-\theta )}^{2}}}$, due to ${{r}^{2}}>>r$, so $r$ is may omitted, the formula becomes ${{r}^{2}}\approx N\ln \frac{1}{{{(1-\theta )}^{2}}}$. Thus we can obtain $r\approx \sqrt{2N\ln \frac{1}{1-\theta }}$.

Furthermore, if selecting $\theta =0.5$, and let $N={{N}_{e}}$, according to the lemma 2.2 and the Deduction 2.1, at the same time, we think that the number of the positive odd in $N$ existing ${{N}_{e}}/2$ number, in fact, $r$ choices are only in the range operation of ${{N}_{e}}/2$. So we put the parameters that $\theta =0.5$ and ${{N}_{e}}/2$ into the formula, finally obtain as follow result.
\[r\approx 0.8\text{33}\sqrt{{{N}_{e}}}\]

The results show that, by random selection and computing the value of the element ${{q}_{j}}$ in $Q$, just need computing the number value of about $r\approx 0.8\text{33}\sqrt{{{N}_{e}}}$, which can find out ${{q}_{j}}$ is a matching prime, and ${{G}_{ij}}^{'}({{N}_{e}})$ satisfy the matching requirement of the Goldbach number $G({{N}_{e}})$. This result has been proved that in $\bmod \overset{\equiv }{\mathop{M}}\,({{N}_{e}})$, we any given one ${{N}_{e}}$, once the prime in $P$ is determined, if  want to find out the prime matching result of element ${{q}_{j}}$ in Q, if only computing the number value of about $r\approx 0.8\text{33}\sqrt{{{N}_{e}}}$ , it can  conform to the matching requirement of the Goldbach number $G({{N}_{e}})$ in the probability of 50\%. This result confirms that element ${{q}_{j}}$ in Q probably choose more than once is a prime number, but at least can assure that  one element ${{q}_{j}}$ is a prime number, and it makes ${{G}_{ij}}^{'}({{N}_{e}})$ satisfy the matching requirement of the Goldbach number $G({{N}_{e}})$; or just only have one prime is selected ( other is odd numbers), but it can conform ${{G}_{ij}}^{'}({{N}_{e}})$ and satisfy the requirements of the Goldbach number $G({{N}_{e}})$.

\textbf{Theorem 3.3.} There is at least one formula that satisfies the matching requirement of Even Goldbach Guess in$\bmod\, \overset{\equiv }{\mathop{M}}\,({{N}_{e}})$.

\textbf{Proof.} Because changing the value of different probability $\theta $ in $\bmod \overset{\equiv }{\mathop{M}}\,({{N}_{e}})$,the element ${{q}_{\text{j}}}$ based on different probabilities maybe appear one prime, and  conforms  ${{G}_{ij}}^{'}({{N}_{e}})$ satisfies the matching requirement of Goldbach number $G({{N}_{e}})$; or the element ${{q}_{\text{j}}}$ may choose more than once is a prime number, but at least can assure that one element ${{q}_{j}}$ is a prime number, and makes ${{G}_{ij}}^{'}({{N}_{e}})$ satisfy the matching requirement of the Goldbach number $G({{N}_{e}})$. About a lower bound of $r$ value of corresponding to the calculated quantity of the element ${{q}_{j}}$ as follow.
\[0.\text{3}25\sqrt{{{N}_{e}}}\le r\le 2.146\sqrt{{{N}_{e}}}\quad  (0.10\le \theta \le 0.99)\]

Because according to the formula of $r\approx \sqrt{{{N}_{e}}\ln (1/(1-\theta ))}$, we can obtain random calculation the how many range of $r$ value numbers by selecting different typical probability $\theta $ values. For examples: when $\theta $ = $0.10$, $r\approx 0.325\sqrt{{{N}_{e}}}$; when $\theta = 0.50, r\approx 0.833\sqrt{{{N}_{e}}}$; when $\theta $ = $0.99$, $r\approx 2.146\sqrt{{{N}_{e}}}$. Obviously, the smaller the value in $\theta $, the smaller the value of $r$; on the contrary, if the larger the value of $\theta $, the larger the value of $r$. Even if the value of $\theta $ is very important, 
but the value of  $ \ln (1/(1-\theta ))$ still smaller,general r is in proportion to $\sqrt{{{N}_{e}}}$. 
Here the parameter value is only used as a reference value, which indicates a possible selection form that the probability ranges from  10\% to 50\%, and again to close 100\%. In fact, when $\theta $ = 100, $r>2.146\sqrt{{{N}_{e}}}$. In $\bmod\, \overset{\equiv }{\mathop{M}}\,({{N}_{e}})$, whether the size of ${{N}_{e}}/2$ and $\theta $ are how changed, it is all instruction a fact, that is to randomly select the computing value of the element ${{q}_{j}}\;(1\le j\le r)$ of corresponding to the prime ${{p}_{i}}\;(1\le i\le r)$, if only by calculate different the range value of $r'\sqrt{{{N}_{e}}}\;(0.325\le r'\le 2.146)$, the result is can determine the element value of ${{q}_{j}}$, and which in different probability calculated quantity to find out corresponding a prime, and it conforms ${{G}_{ij}}^{'}({{N}_{e}})$ to satisfy the requirements of the Goldbach number $G({{N}_{e}})$; Or when the probability $\theta =100$, the element ${{q}_{j}}$ if only also computing the value of $r>2.146\sqrt{{{N}_{e}}}$, the result is can be find out a determine prime and makes ${{G}_{ij}}^{'}({{N}_{e}})$ to satisfy the requirements of the Goldbach number $G({{N}_{e}})$. Thus will ensure that there is at least one formula to satisfy the result of even Goldbach Guess in $\bmod \overset{\equiv }{\mathop{M}}\,({{N}_{e}})$.

As a comparison reference between the calculated values and the actual matching values, the data illustrates a different set of $\theta/r$ values and the actual matching values of the even Goldbach Guess are shown in Table 1 below.

Where, $r$ value is the range value of $r'\sqrt{{{N}_{e}}}$ calculated for random calculation when the corresponding typical probability $\theta $ = 0.2, $\theta $ = 0.5 and $\theta $ = 0.99 is obtained. The result illustration the element ${{q}_{j}}$ in $Q$ can ensure that at least have a prime which satisfy the matching requirements of Goldbach number $G({{N}_{e}})$. While at the table 1, the actual value is to refers when different${{N}_{e}}, \varphi (G({{N}_{e}}))$ is the matching number of real has been existed even Goldbach's numbers. For instance, when ${{N}_{e}}$ = 560000, the probability is $\theta $ = 0.2, $\theta $ = 0.5, $\theta $ = 0.99, the element ${{q}_{j}}$ in $Q$, by randomly calculating the value of $r'\sqrt{{{N}_{e}}}$, it obtains the value  of corresponding each computing result is $r$ = 353, $r$ = 623 and $r$ = 1606. The result also illustrates it can ensure have one prime may satisfy the requirement of $G({{N}_{e}})$. And real value is mean when ${{N}_{e}}$ = 560000, it really existing $\varphi (G({{N}_{e}}))$ = 3971. Other parameters $\theta /r$ and so on. Above these result further show that there is at least one prime can satisfy the matching requirement of the Goldbach number $G({{N}_{e}})$.

\centerline{Table 1 the reference parameters of $\theta /r$ and the actual matching values}
 \[\left\{ {\begin{array}{*{20}{c}}
{{N_e}}&{\begin{array}{*{20}{c}}
{\theta  = 0.2}\\
r
\end{array}}&{\begin{array}{*{20}{c}}
{\theta  = 0.5}\\
r
\end{array}}&{\begin{array}{*{20}{c}}
{\theta  = 0.99}\\
r
\end{array}}&\begin{array}{l}
{\mathop{\rm Re}\nolimits} al\\
value
\end{array}\\
{100}&5&8&{21}&6\\
{2688}&{24}&{43}&{111}&{88}\\
{6000}&{37}&{65}&{166}&{178}\\
{30000}&{82}&{144}&{372}&{602}\\
{60000}&{116}&{204}&{526}&{1084}\\
{100000}&{149}&{263}&{679}&{810}\\
{300000}&{259}&{456}&{1175}&{3915}\\
{560000}&{353}&{623}&{1606}&{3971}\\
{3000000}&{818}&{1443}&{3717}&{27502}\\
{60000000}&{3656}&{6452}&{16623}&{371226}\\
{1000000000}&{14926}&{26342}&{67862}&{2274205}\\
{...}&{...}&{...}&{...}&{...}
\end{array}} \right.\]

This part of the probabilistic proof show that, the basic judgement of existed probability of even Goldbach conjecture $G({{N}_{e}})$ is always greater than the probability of non-existence. In fact, in the model $G{{N}_{e}}TM$, the result of the controller calculation shows the correct matching state, and the basic facts are all true (T). And the result of the controller calculation seems to be incorrect matching state is false (F), which is only a very small probability. This means that the possibility of the $G{{N}_{e}}TM$ model halting problem is very low and almost impossible.

\section{Computer recursion solvable proving on the prime matching rule algorithm in controller \protect\footnote{The content of the section 4 is to reference as the footnote illustration of the section 3.}}

What is the meaning of the prime number matching rule algorithm in the model $G{{N}_{e}}TM$ controller, which refers to the design of a computer recursive solvable algorithm to satisfy the prime number match of even Goldbach conjecture through machine operation.

This section will main discuss fundamental problems of computer recursion that can be solved by the prime number matching rule algorithm in the model $G{{N}_{e}}TM$ controller, as well as constructing the model basis, recursive structural feature descriptions and the logic criterion model of even Goldbach conjecture existence, all these major analyses and general models will also give proof.

\subsection{some main lemmas}

According to definition 2.3, in the independent closure of ${{N}_{e}}$, it's know that $Mod\,\overset{\equiv }{\mathop{X}}\,(o)$ can be understood as a combination of odd matching, the result is an implicit described set that full permutation of two element add sums consists of the congruence relations. That is to say, all the expressions that have established the congruence sum relationship include the full  permutation of the odd positive integers of $x$ and $y$. The explicit described set is the set of complete permutation of a kind of same even number function ${{\overset{{}}{\mathop{{{f}_{\sum }}(N}}\,}_{e}})={{f}_{\Sigma }}({{x}_{i}},{{y}_{j}})=$ $\{{{N}_{(1,2n-1)}},$ ${{N}_{(3,2n-3)}},\cdots ,$ ${{N}_{(2n-3,3)}}$, ${{N}_{(2n-1.1)}}\}$, it is written as: $\bmod\, \overset{\equiv }{\mathop{X}}\,{{(o)}^{*}}$. Another way to express it by the set, that is:

\textbf{Definition 4.1.} We call the set of $\bmod \, \overset{\equiv }{\mathop{X}}\,{{(o)}^{*}}$ is a strictly primitive recursion enumerable described set. Especially it refer to in the independent closure of ${{N}_{e}}$, if there's one recursive functions $f({{x}_{i}},{{y}_{j}})={{x}_{i}}+{{y}_{j}}\equiv {{N}_{e}}$ of full permutation result, which makes ${{f}_{\sum }}({{x}_{i}},{{y}_{j}})$ = $N_{e}^{\Sigma }$ $(1\le i,j\le 2n-1)$ ($i,j$ are may be exchanged repeat) of ${{N}_{e}}/2$ even numbers are all established full permutation existence result of two number add sum relation.

\textbf{Lemma 4.1.} $\bmod\, \overset{\equiv }{\mathop{X}}\,(o)$ is also a described set of strictly primitive recursion enumerable.

\textbf{ Proof.} According to the definition of $Mod\,\overset{\equiv }{\mathop{X}}\,{{(o)}^{*}}$ and definition 2.3, because of $f:Mod\,\overset{\equiv }{\mathop{X}}\,{{(o)}^{*}}$ $\to Mod\,\overset{\equiv }{\mathop{X}}\,(o)$, thus two forms is equivalence, i.e.$Mod\,\overset{\equiv }{\mathop{X}}\,{{(o)}^{*}}$ $\cong $ $Mod\,\overset{\equiv }{\mathop{X}}\,(o)$, then has:$\{{{N}_{e1}}=({{x}_{1}},y{}_{2n-1}),$ ${{N}_{e3}}=({{x}_{3}},{{y}_{2n-3}})$,$\cdots ,$ ${{N}_{e(2n-1)}}=({{x}_{2n-1}},{{y}_{1}})\}$ $\cong $ $\{{{N}_{e1}}\equiv {{y}_{2n-1}}(\bmod\, {{x}_{1}}),$
${{N}_{e3}}\equiv {{y}_{2n-3}}(\bmod \,{{x}_{3}}),\cdots ,{{N}_{e(2n-1)}}\equiv {{y}_{1}}(\bmod \,{{x}_{2n-1}})\}$; Similarly, $f: Mod\,\overset{\equiv }{\mathop{X}}\,(o)\to $ $Mod\overset{\equiv }{\mathop{X}}\,{{(o)}^{*}}$,  the two forms are equivalent too: $Mod\,\overset{\equiv }{\mathop{X}}\,(o)\cong $ $Mod\,\overset{\equiv }{\mathop{X}}\,{{(o)}^{*}}$. For in the independent closure of ${{N}_{e}}$, ${{N}_{e}}/2$ even $N_{e}^{\Sigma }$ established the even full permutation all can be strictly enumerated in random, and which can consist a group of the complete add sum forms, that result is all the odd number combination permutation.

\textbf{Lemma 4.2.} The set of $Mod\,\overset{\equiv }{\mathop{X}}\,{{(o)}^{*}}$ belongs to the primitive recursion set.

\textbf{Proof.} According to the definition of the primitive recursion set, if $Mod\,\overset{\equiv }{\mathop{X}}\,{{(o)}^{*}}$ is a primitive recursion set, if and only if the individual characteristic function of $Mod\,\overset{\equiv }{\mathop{X}}\,{{(o)}^{*}}$ is a recursive function. Might as well make ${{A}^{*}}=$ $Mod\,\overset{\equiv }{\mathop{X}}\,{{(o)}^{*}}$, then there is:\\
\begin{equation}
 \nonumber
{{C}_{{{A}^{*}}(individual)}}=({{x}_{i}},{{y}_{j}})=\left\{ \begin{matrix}
   1,\begin{matrix}
   {} & {}  \\
\end{matrix}f({{x}_{i}},{{y}_{j}})=(x{}_{i}+{{y}_{j}})={{N}_{e}}\in {{A}^{*}}  \\
   0,\begin{matrix}
   {} & {}  \\
\end{matrix}f({{x}_{i}},{{y}_{j}})=(x{}_{i}+{{y}_{j}})={{N}_{e}}\notin {{A}^{*}}  \\
\end{matrix} \right.
\end{equation}

Because ${{A}^{*}}$ is the set of the function value range of $f({{x}_{i}},{{y}_{j}})\equiv {{N}_{e}}$, that is:
\[{{A}^{*}}=\{<{{x}_{i}},{{y}_{j}}>|:f({{x}_{i}},{{y}_{j}})=({{x}_{i}}+{{y}_{j}})\equiv {{N}_{e}}^{{}};i,j=1,3,\ldots ,{{N}_{e}}-1\}\]

So, the general characteristic function of ${{A}^{*}}$ also can be represented as:
\[{{C}_{{{A}^{*}}}}(<{{x}_{i}},{{y}_{j}}>|:i,j=1,3,\ldots ,{{N}_{e}}-1)={{\bar{s}}_{g}}|(f({{x}_{i}},{{y}_{j}})|:i,j=1,3,\ldots ,{{N}_{e}}-1)-{{N}_{e}}|\]

And because $f({{x}_{i}},{{y}_{j}})$ is a strictly original recursion enumerated function within the even number ${{N}_{e}}$, so ${{A}^{*}}$ = $Mod\,\overset{\equiv }{\mathop{X}}\,{{(o)}^{*}}$ is belong to the primitive recursion set.

\textbf{Deduction 4.1.} The set of $Mod\,\overset{\equiv }{\mathop{X}}\,(o)$ belongs to the primitive recursion set too.

\textbf{Proof.} Because of $Mod\,\overset{\equiv }{\mathop{X}}\,{{(o)}^{*}}$ $\cong $ $Mod\,\overset{\equiv }{\mathop{X}}\,(o)$, according to definition 2.3 and 4.1, as well as lemma 2.2, it's easy to prove that the deduction is established.

\textbf{Lemma 4.3.} The following basic set of functions are all primitive recursion functions:

(1)$f(x,y)=x+y$            \qquad\,  (2)$f(x,y)=x\dot{-}y$

(3)$f(x,y)=|x-y|$            \qquad(4)\,$div(x,y)$

(5) ${{p}_{r}}(x)=\left\{ \begin{matrix}
   1,\begin{matrix}
   {} & {}  \\
\end{matrix}x-is{{}_{{}}}\, prime  \\
   ^{{}}0,\begin{matrix}
   {} & else\begin{matrix}
   \begin{matrix}
   {} & {}  \\
\end{matrix} & {} & {}  \\
\end{matrix}  \\
\end{matrix}  \\
\end{matrix} \right.$

(6) ${{p}_{x}}$ is the $x$ order prime number (Appoint ${{p}_{0}}=0$)

\textbf{Proof.} According to the concept of the primitive recursion function and the property of the basic definition, it can be easily proved the lemma is true.

\textbf{Lemma 4.4.} ${{N}_{e}}\equiv {{y}^{{}}}(\bmod\, x)$ is a primitive recursive predicate.

\textbf{Proof.} According to the primitive recursion predicate, it is obvious that the congruence expression to ${{N}_{e}}$ $\equiv {{y}^{{}}}(\bmod \,x)$ is a primitive recursion predicate. If the congruence relationship expression of ${{N}_{e}}\equiv {{y}^{{}}}(\bmod \,x)$ is the primitive recursion predicate.In the independent closure of ${{N}_{e}}$,${{N}_{e}}\equiv {{y}^{{}}}(\bmod \, x)$ is a given determined expression, which is one to one correspond to the expression form of the element add summation, i.e.,
\[{{N}_{e}}\equiv {{y}^{{}}}(\bmod\, x)\Leftrightarrow {{N}_{e}}=x+y\]

Of which: $x,y\in {{N}_{e}}$, and $(x+y)/{{N}_{e}}\equiv 1$. Here Might as well make the predicate:
\[P(x,y,{{N}_{e}})=[{{N}_{e}}\equiv {{y}^{{}}}(\bmod\, x)]\]

Because the $P(x,y,{{N}_{e}})$ is a ternary predicate, the characteristic function of P is:
\[{{C}_{p}}(x,y,{{N}_{e}})=\left\{ \begin{matrix}
   ^{{}}1,\begin{matrix}
   {} & {}  \\
\end{matrix}P(x,y,{{N}_{e}})-true  \\
   0,\begin{matrix}
   {} & {}  \\
\end{matrix}else\begin{matrix}
   \begin{matrix}
   {} & {}  \\
\end{matrix} & {} & {}  \\
\end{matrix}  \\
\end{matrix} \right.\]

And because when the characteristic function is true, its equivalence relationship is as follow.
\[{{C}_{p}}(x,y,{{N}_{e}})=1\cong div((x,y),{{N}_{e}})\]

Due to $div((x,y),{{N}_{e}})$ is a primitive recursive function, ${{C}_{p}}(x,y,{{N}_{e}})$ is a primitive recursion function too. So,$P(x,y,{{N}_{e}})$ is a primitive recursion predicate. Thus, the congruence relationship expression of ${{N}_{e}}\equiv {{y}^{{}}}(\bmod \,x)$ is a primitive recursion predicate.

\textbf{Lemma 4.5.} $Mod\,\overset{\equiv }{\mathop{X}}\,(o)$ = (${{N}_{e}}{{(o)}^{{}}}$ ${{\underline{{\bar{\equiv }}}}_{{}}}$ ${{Y}_{o}}^{{}}(\bmod\, {{X}_{o}})$) is an enumerable strict primitive recursive predicate.

\textbf{Proof.} Suppose the predicate is $P(X,Y,{{N}_{e}})=$ $({{N}_{e}}{{(o)}^{{}}}$ ${{\underline{{\bar{\equiv }}}}_{{}}}$ ${{Y}_{o}}^{{}}(\bmod \,{{X}_{o}}))$, consider that there is always an enumerator of $k$ acting on the variables of $x$ and $y$ in the independent closure of ${{N}_{e}}$, which makes it expand or shrink. So there is:
\[f:P(X,Y,{{N}_{e}})\to \exists {{k}^{{}}}[(x\underline{+}k)+(y\overline{+}k)\equiv {{N}_{e}}]\Rightarrow \exists k[(x\underline{+}k)\;{{\equiv }_{{ \;}}}(y\bar{+}k)(\bmod\, {{N}_{e}}(o))]\]

Obviously, if $P(X,Y,{{N}_{e}})$ is a strictly primitive recursion enumerable predicate, if and only if when there is a full defined primitive recursion predicate $Q(x,y,{{N}_{e}},k)$ in independent closure of given ${{N}_{e}}$, the following relationship is established:
\[P(X,Y,{{N}_{e}})\leftrightarrow (\exists k)Q(x,y,{{N}_{e}},k)\]

Additionally, considering that $Q(x,y,{{N}_{e}},k)$ in the closed $Mod\,\overset{\equiv }{\mathop{X}}\,(o)$, which the primitive recursion predicate of the matching definition of existing corresponding parameters, and $P(X,Y,{{N}_{e}})$$\leftrightarrow $$(\exists k)Q(x,y,{{N}_{e}},k)$ is established. If given any ${{N}_{e}}$, then there is:
\[f(x,y,k)=\left\{ \begin{matrix}
   {{f}_{1}}(x,y,k),\begin{matrix}
   {} & {}  \\
\end{matrix}(x,y,k)\in {{N}_{1}}\equiv {{N}_{e}}  \\
   {{f}_{3}}(x,y,k),\begin{matrix}
   {} & {}  \\
\end{matrix}(x,y,k)\in {{N}_{3}}\equiv {{N}_{e}}  \\
   \vdots   \\
   {{f}_{2n-1}}(x,y,k),\begin{matrix}
   {} & (x,y,k)\in {{N}_{2n-1}}\equiv {{N}_{e}}  \\
\end{matrix}  \\
   non-definning,\begin{matrix}
   {} & other\begin{matrix}
   \begin{matrix}
   {} & {} & {}  \\
\end{matrix} & {} & {}  \\
\end{matrix}  \\
\end{matrix}  \\
\end{matrix} \right.\]

Obviously, $f(x,y,k)$ is a primitive recursion function, it comprises of the set of the congruence matching domain and value range for corresponding parameters and matching definition, and it is strictly primitive recursion can be enumerated in the closed ${{N}_{e}}$. So
\[P(X,Y,{{N}_{e}}):Mod\,\overset{\equiv }{\mathop{X}}\,(o)={{N}_{e}}{{(o)}^{{\;}}}{{\underline{{\bar{\equiv }}}}_{{\;}}}{{Y}_{o}}^{{}}(\bmod\, {{X}_{o}})\]
is a strictly primitive recursion enumerable predicate.

\textbf{Lemma 4.6.} If ${{x}_{i}}\in X\in {{N}_{e}},{{y}_{j}}\in Y\in {{N}_{e}}$, and ${{x}_{i}}\ne {{y}_{j}}$ $(or:{{x}_{i}}={{y}_{j}})$. Suppose that the predicate $P({{x}_{i}})$ means ${{x}_{i}}$ is the prime of the location of $X$, and the predicate $P({{y}_{j}})$ means that ${{y}_{j}}$ is the prime of the location of $Y$. Then, the following function
\[{{P}_{r}}({{x}_{i}})(or:{{P}_{r}}({{y}_{j}}))=\left\{ \begin{matrix}
   1,\begin{matrix}
   {} & {}  \\
\end{matrix}{{x}_{i}}(or:{{y}_{j}})-is\,prime  \\
   0,\begin{matrix}
   {} & {} & else\begin{matrix}
   {} & {} & {} & {} & {}  \\
\end{matrix}  \\
\end{matrix}  \\
\end{matrix} \right.\]
is a primitive recursion function, and both of the $P({{x}_{i}})$ and $P({{y}_{j}})$ are primitive recursion predicates.

\textbf{Proof.} According to the function of (5) and (6) in lemma 4.3, it's easy to prove that the lemma is true.

Additionally, the lemma indicates that the discriminant of the prime numbers can be computed recursively in the independent closure of ${{N}_{e}}$.

\textbf{Lemma 4.7.} Suppose that $P({{x}_{i}})$, $P({{y}_{j}})$, $P({{x}_{i}},{{y}_{j}},{{N}_{e}})$ are primitive recursion predicates, then $P({{x}_{i}})$ $\cap $ $P({{y}_{j}})$ $\cap $ $P({{x}_{i}},{{y}_{j}},{{N}_{e}})$ is a primitive recursion predicate, too.

\textbf{Proof.} According to lemma 4.3, lemma 4.6, and the property of characteristic function of $P({{x}_{i}})$ and $P({{y}_{j}})$, the proof of the lemma is easy.
\subsection{The logic judge model of even Goldbach conjecture existing}

\textbf{Definition 4.2.} If even Goldbach conjecture can be described as following model[12-13]:

(1)$\,\forall nP(n)$

(2)$\,(\forall {{N}_{e}}){{(\exists x{}_{i})}_{<{{N}_{e}}}}{{(\exists {{y}_{j}})}_{<{{N}_{e}}}}T({{x}_{i}},{{y}_{j}},{{N}_{e}})$

Then the formula (1) is called general form, the expression of $\forall n$ can make $p(n)$ become true; the formula (2) is called specifically representation form, it means that there are elements ${{x}_{i}}$ and ${{y}_{j}}$ that make $T({{x}_{i}},{{y}_{j}},{{N}_{e}})$ true, and each even number ${{N}_{e}}$ are established. $T({{x}_{i}},{{y}_{j}},{{N}_{e}})$ means that the add sum of ${{x}_{i}}$ and ${{y}_{j}}$ are equal to ${{N}_{e}}$. $P(n)={{(\exists {{x}_{i}})}_{<{{N}_{e}}}}{{(\exists {{y}_{j}})}_{<{{N}_{e}}}}T({{x}_{i}},{{y}_{j}},{{N}_{e}})$ is called the matching predicate of even Goldbach conjecture.

So, we can easily get the following lemma.

\textbf{Lemma 4.8.} The bound existential quantifiers of ${{(\exists {{x}_{i}})}_{<{{N}_{e}}}}$ and ${{(\exists {{y}_{j}})}_{<{{N}_{e}}}}$ are primitive recursion.

\textbf{Proof.} The element $i$ and $j$ for in ${{(\exists {{x}_{i}})}_{<{{N}_{e}}}}$ and ${{(\exists {{y}_{j}})}_{<{{N}_{e}}}}$, as long as both satisfy the matching requirement of given even ${{N}_{e}}$, and to conform to the element range of enumerating choice $1\le i,j\le 2n-1=N{}_{e}-1$, then below the operations of ${{(\exists {{x}_{i}})}_{<{{N}_{e}}}}$ and ${{(\exists {{y}_{j}})}_{<{{N}_{e}}}}$, which are all primitive recursion.

\textbf{Lemma 4.9.} The matching predicate of even Goldbach conjecture is independently closed in the operation of the bound existential quantifiers.

\textbf{Proof.}  If $T({{x}_{i}},{{y}_{j}},{{N}_{e}})$ is true,then ${{N}_{e}}{{(o)}^{{\;}}}{{\underline{{\bar{\equiv }}}}_{{\;}}}{{Y}_{o}}^{{}}(\bmod\, {{X}_{o}})$ is true, too. The matching relationship of even Goldbach conjecture for $\forall {{N}_{e}}\ge 6$, it certainly exists the odd complete congruence expressions of independently closed even ${{N}_{e}}$. In order to find out the specific element ${{x}_{i}},{{y}_{j}}$ in $Mod\,\overset{\equiv }{\mathop{X}}\,(o)$, it must be satisfied with the following corresponding relations:
    \[T({{x}_{i}},{{y}_{j}},{{N}_{e}})=1 \vdash {{(\exists {{x}_{i}})}_{<{{N}_{e}}}}{{(\exists {{y}_{j}})}_{<{{N}_{e}}}}T({{x}_{i}},{{y}_{j}},{{N}_{e}})\]
It is always true. Obviously, Here the elements ${{x}_{i}}^{{}}an{{d}^{{}}}{{y}_{j}}$ are all bounded. i.e., $1\le {{x}_{i}}\in X$ $\le 2n-1={{N}_{e}}-1$,$1\le {{y}_{j}}\in Y\le 2n-1=N{}_{e}-1$, The process of enumerating and checking the elements ${{x}_{i}}$ and ${{y}_{j}}$ must ensure that :
        \[T({{x}_{i}},{{y}_{j}},{{N}_{e}})=1\leftrightarrow {{(\exists {{x}_{i}})}_{\le {{N}_{e}}-1}}{{(\exists {{y}_{j}})}_{\le {{N}_{e}}-1}}T({{x}_{i}},{{y}_{j}},{{N}_{e}})=1\]
is established. Thus, for $\forall {{N}_{e}}\ge 6$, as long as the given even number ${{N}_{e}}$ is determined by the enumeration of each odd element obtained, the enumeration can definitely make:
                      \[{{N}_{e}}{{(o)}^{{}}}{{\underline{{\bar{\equiv }}}}_{{}}}{{Y}_{o}}^{{}}(\bmod \, {{X}_{o}})\]
is true. So have
                 \[P(n)={{(\exists {{x}_{i}})}_{\le {{N}_{e}}-1}}{{(\exists {{y}_{j}})}_{\le {{N}_{e}}-1}}T({{x}_{i}},{{y}_{j}},{{N}_{e}})\]

And the operations of the predicate relationship are all independent closure operations within given ${{N}_{e}}$. That is, the operations of the inbound existential quantifier is also closed independently. In fact, enumerate the operation of each element ${{x}_{i}},{{y}_{j}}$ whether the prime computing in $Mod\,\overset{\equiv }{\mathop{X}}\,(o)$ is closed independently, too.

\textbf{Lemma 4.10.} The matching predicate of even Goldbach conjecture corresponding under the operation of the universal quantifiers, as long as the given ${{N}_{e}}$ is arbitrarily enumerable, then  for $\forall ({{N}_{e}}\ge 6)\;({{N}_{e}}=2n+4,n\ge 1{{,}_{{}}}n-Given)$ operations are closed independently.

\textbf{Proof.} Exploring the universal quantifiers $(\forall {{N}_{e}}\ge 6)$, let ${{N}_{e}}=\{(2n+4),$ $(n\ge 1)\}$, if $n$ is given enumerable, then ${{N}_{e}}$ is given, too. When $n\to {{\infty }^{-}}$,${{N}_{e}}$ is a given large enough even number, it is written as $N_{e}^{{{\infty }^{-}}}$, and ${{N}_{e}}+1$ is written as $N_{e}^{{{\infty }^{+}}}$, and $N_{e}^{{{\infty }^{-}}}<N_{e}^{{{\infty }^{+}}}$. When ${{N}_{e}}=\{(2n+4),(n\ge 1)\}$ $\le N_{e}^{{{\infty }^{-}}}$, the selecting $n$=1,2,$\ldots ,{{\infty }^{-}}$, then there is:
      \[Mod\,\overset{\equiv }{\mathop{{{X}_{6}}}}\,(o)<Mod\,\overset{\equiv }{\mathop{{{X}_{8}}}}\,(o)<\cdots <\underset{n\to {{\infty }^{-}}}{\mathop{Mod}}\,{{\overset{\equiv }{\mathop{X}}\,}_{2n+4}}(o)=Mod\,\overset{\equiv }{\mathop{X_{e}^{{{\infty }^{-}}}}}\,(o)\]

Given a large enough even number $N_{e}^{{{\infty }^{-}}}$, the even ${{N}_{e}}$ for $\forall \{2n+{{4}^{{}}}(n\ge 1)\}\le N_{e}^{{{\infty }^{-}}}$ in $Mod\,\overset{\equiv }{\mathop{X}}\,(o)$, there is always a given even ${{N}_{e}}$ corresponding to one of it. As long as $N_{e}^{{{\infty }^{-}}}<N_{e}^{{{\infty }^{+}}}$, which corresponding to full permutation of the even is sure to cover the even numbers of $\forall ({{N}_{e}}\ge 6)<N_{e}^{{{\infty }^{-}}}$. In this way, for $(\forall {{N}_{e}}\ge 6)$, within any enumerated or given ${{N}_{e}}<{{N}_{e}}+1$ the operation of the corresponding universal quantifiers, whatever enumerating ${{x}_{i}}$ and ${{y}_{j}}$, or discussing the operation of the matching predicate of even Goldbach conjecture about $P(n)={{(\exists {{x}_{i}})}_{\le {{N}_{e}}-1}}{{(\exists {{y}_{j}})}_{\le {{N}_{e}}-1}}T({{x}_{i}},{{y}_{j}},{{N}_{e}})$, as long as $Mod\,\overset{\equiv }{\mathop{{{X}_{i}}}}\,(o)$$(6\le i\le {{\infty }^{-}})$ $\le Mod\,\overset{\equiv }{\mathop{X_{e}^{{{\infty }^{-}}}}}\,(o)$) , the element ${{x}_{i}}^{{}}an{{d}^{{}}}{{y}_{j}}$(including it oneself is prime numbers) within corresponding ${{N}_{e}}$, every operation form is all independently closed.
\subsection{ Main judge results}

In order to more clearly express even Goldbach conjecture about the prime matching rule algorithm in the new model $G{{N}_{e}}TM$, and the machine computing is recursively solvable, we must further judge $p(n)$. According to above  definition and  lemma, the next the judgment result of computer recursion calculation of even Gothbach conjecture is given.

\textbf{Theorem 4.1.} The matching predicate of even Goldbach conjecture is primitive recursion predicate.

\textbf{Proof.} Given the ${{N}_{e}}$, those prime number are not difficult to determine in $Mod\,\overset{\equiv }{\mathop{X}}\,(o)$. According to the functions(5) and (6) in the lemma 4.3, ${{p}_{r}}(x)$ and ${{p}_{x}}$ are primitive recursive functions. In order to make sure the prime matching of even Goldbach number in $Mod\,\overset{\equiv }{\mathop{X}}\,(o)$, Might as well make
        \[P(n)={{(\exists {{x}_{i}})}_{<}}_{{{N}_{e}}}{{(\exists {{y}_{j}})}_{<{{N}_{e}}}}T({{x}_{i}},{{y}_{j}},{{N}_{e}})\]

If $P(n)=1$, if and only if there must be $P(n)=1$ $\leftrightarrow $$T({{x}_{i}},{{y}_{j}},{{N}_{e}})=1$. (the described relationship is true).

So, as long as to prove that $T({{x}_{i}},{{y}_{j}},{{N}_{e}})$ is a primitive recursion predicate, then $P(n)$ is a primitive recursion, too. Because according to lemma 4.8, we can know that
        \[T({{x}_{i}},{{y}_{j}},{{N}_{e}})\leftrightarrow {{N}_{e}}{{(o)}^{{\;}}}{ }{{\underline{{\bar{\equiv }}}}_{{\;}}}{ }{{Y}_{o}}^{{}}(\bmod\, {{X}_{o}})\]
is established. More specifically, here let
          \[T({{x}_{i}},{{y}_{j}},{{N}_{e}})=(P({{x}_{i}})\cap P({{y}_{j}})\cap ({{x}_{i}}+{{y}_{j}}=2n+4={{N}_{e}})),\]
if $T({{x}_{i}},{{y}_{j}},{{N}_{e}})=1$$\Rightarrow $$(P({{x}_{i}})\cap P({{y}_{j}})\cap ({{x}_{i}}+{{y}_{j}}={{N}_{e}}))$ = 1

From the lemma 4.6 known that both ${{P}_{r}}({{x}_{i}})$ and ${{P}_{r}}({{y}_{j}})$ are primitive recursion functions, $f({{x}_{i}},{{y}_{j}})={{x}_{i}}+{{y}_{j}}$ is a primitive recursion function, too. And $P({{x}_{i}})$, $P({{y}_{j}})$, $({{x}_{i}}+{{y}_{j}}\equiv {{N}_{e}})$ are primitive recursion predicates. And the characteristic function described in terms of ${{C}_{T}}$(${{x}_{i}},$ ${{y}_{j}},$ ${{N}_{e}})$, $T({{x}_{i}}$, ${{y}_{j}},$ ${{N}_{e}})$ is a primitive recursion function,too. So, $T({{x}_{i}},$ ${{y}_{j}},$ ${{N}_{e}})$ is a primitive recursion predicate. Thus, in the independent closure of the even number ${{N}_{e}}$, whether given or enumerated any ${{N}_{e}}$, The predicates of judgment described by:
         \[P(n)={{(\exists {{x}_{i}})}_{<}}_{{{N}_{e}}}{{(\exists {{y}_{j}})}_{<{{N}_{e}}}}T({{x}_{i}},{{y}_{j}},{{N}_{e}})\]
that is a primitive recursion predicate, too.

Additionally, according to the definition 4.1, the lemma 4.1 and the lemma 4.5, if  consider another situation, suppose that ${{T}^{*}}({{x}_{i}},{{y}_{j}},{{N}_{e}})$ = {$f(0),$ $f(3),\ldots ,f(2n-1)$ is a recursive enumerable set, when enumerating every ${{f}_{l}}\;(l=1,3,\ldots ,2n-1)$ which can all make
        \[{{N}_{e}}{{(o)}^{{}}}{\;}{{\underline{{\bar{\equiv }}}}_{{}}}{\;}{{Y}_{o}}^{{}}(\bmod\, {{X}_{o}})\]
Corresponding establish, or so to speak, as follow the predicate,
                \[P(n)={{(\exists {{x}_{i}})}_{<}}_{{{N}_{e}}}{{(\exists {{y}_{j}})}_{<{{N}_{e}}}}T({{x}_{i}},{{y}_{j}},{{N}_{e}})\]
is a predicate of strictly closed primitive recursion enumerable, too.

\textbf{Theorem 4.2.} The judgment problem of even Goldbach conjecture existence is the computer recursion solvable.

\textbf{Proof.} Supposing any even ${{N}_{e}}$ of even Goldbach conjecture exist the model $Mod\,\overset{\equiv }{\mathop{X}}\,(p)$, for $Mod\,\overset{\equiv }{\mathop{X}}\,(p)$ $\in $ $Mod\,\overset{\equiv }{\mathop{X}}\,(o)$, i.e, There is an implicit expression:
            \[{{N}_{e}}{{(o)}^{{}}}{\;}{{\underline{{\bar{\equiv }}}}_{{\;}}}{{Y}_{o}}^{{}}(\bmod\, {{X}_{o}})\]
And every explicit expression of even ${{N}_{e}}$ as follow:
              \[{{x}_{i}}+{{y}_{j}}\equiv {{N}_{e}}, ({{x}_{i}}\in X,\;{{y}_{j}}\in Y{{;}_{{\;}}}X,Y\in {{N}_{e}})\]

Therefore, we must further discuss all even number of ${{N}_{e}}$ $\ge 6$ whether the problem of computer recursively solvable within limited steps. That is to say, by enumerating (or searching) all prime in X and Y, to determine whether can satisfy the element ${{x}_{i}},{{y}_{j}}$ about all two element add sum can established the congruence relation expressions existing within independent closure of ${{N}_{e}}$. Actually, for $\forall {{N}_{e}}\ge 6$, as long as make a further inductive argument for $P(n)$. For the existence of even Goldbach conjecture can be described synthetically as follow.
\[{{(\forall {{N}_{e}}\ge 6)}_{{{N}_{e}}-any-given}}{{(\exists {{x}_{i}})}_{\le {{N}_{e}}-1}}{{(\exists {{y}_{j}})}_{\le {{N}_{e}}-1}}[P({{x}_{i}})\cap P({{y}_{j}})\cap ({{x}_{i}}+{{y}_{j}}=2n+4\equiv {{N}_{e}})]\]
where, the predicate expression as follow.
    \[p(n)={{(\exists {{x}_{i}})}_{\le {{N}_{e}}-1}}{{(\exists {{y}_{j}})}_{\le {{N}_{e}}-1}}[P({{x}_{i}})\cap P({{y}_{j}})\cap ({{x}_{i}}+{{y}_{j}}=2n+4\equiv {{N}_{e}})]\]

According to the lemma 4.6 ,we know that both ${{P}_{r}}({{x}_{i}})$ and ${{P}_{r}}({{y}_{j}})$ are primitive recursion functions, and both $P({{x}_{i}})$ and $P({{y}_{j}})$ are primitive recursion predicates. Other according to the lemma 4.7 and the theorem 4.1, we know that the predicate of $T({{x}_{i}},{{y}_{j}},{{N}_{e}})$ =
$(P({{x}_{i}})\cap P({{y}_{j}})\cap({{x}_{i}}+{{y}_{j}}=2n+4={{N}_{e}}))$ is a primitive recursion predicate. At the same time, according to the lemma 4.9 and 4.10, the matching predicate of  even Goldbach conjecture as follow.
      \[p(n)={{(\exists {{x}_{i}})}_{\le {{N}_{e}}-1}}{{(\exists {{y}_{j}})}_{\le {{N}_{e}}-1}}[P({{x}_{i}})\cap P({{y}_{j}})\cap ({{x}_{i}}+{{y}_{j}}=2n+4\equiv {{N}_{e}})]\]
%
%

$P(n)$ is primitive recursion predicate in the independent closure of ${{N}_{e}}$. Or we can say that, for $\forall {{N}_{e}}\ge 6$, $P(n)$ corresponding to ${{N}_{e}}$ is strictly primitive recursion enumerable under the situation of bound existential quantifiers or universal quantifiers (the condition of the arbitrary given ${{N}_{e}}$). When ${{N}_{e}}$ is given, the characteristic function of $P(n)$ is a primitive recursion function. Finally, the result of obtained judge must be every even ${{N}_{e}}$ exist as follow relation:
                     \[f:Mod\,\overset{\equiv }{\mathop{X}}\,(o)\to Mod\,\overset{\equiv }{\mathop{X}}\,(p)\]
That is to say, the result exist $Mod\,\overset{\equiv }{\mathop{X}}\,(p)$ = $({{N}_{e}}{{(p)}^{{\;}}}{{\underline{{\bar{\equiv }}}}_{{\;}}}Y{{_{p}^{{}}}^{{}}}(\bmod\, X_{p}^{{}}))$.

Furthermore, we let universal quantifier make a transformation, its operation is confined within independent closure of ${{N}_{e}}$, then the model is transformed into the more intuitively operational judge form:
\[{{(\forall {{x}_{i}})}_{\le {{N}_{e}}-1}}{{(\exists {{N}_{e}}\ge 6)}_{any-given}}{{(\exists {{y}_{j}})}_{\le {{N}_{e}}-1}}[P({{N}_{e}}-{{y}_{j}})\cap P({{y}_{j}})\cap (({{N}_{e}}-{{y}_{j}})+{{y}_{j}}\equiv {{N}_{e}})]\]

Here, the predicate as follow:
     \[p(n)={{(\exists {{N}_{e}}\ge 6)}_{any-given}}{{(\exists {{y}_{j}})}_{\le {{N}_{e}}-1}}[p({{N}_{e}}-{{y}_{j}})\cap p({{y}_{j}})\cap ((N{}_{e}-{{y}_{j}})+{{y}_{j}}\equiv {{N}_{e}})]\]

Similarly, the matching predicate of $p(n)$ for even Goldbach conjecture is also primitive recursion in independent closure of ${{N}_{e}}$. Or we can say that, for $\forall {{N}_{e}}\ge 6$, the predicate                                                                                                       $P(n)$ corresponding to ${{(\forall x{}_{i})}_{\le {{N}_{e}}-1}}$ are all strictly primitive recursion enumerated no matter under the situation of bound existential quantifiers or universal quantifiers (the condition of the arbitrary given ${{N}_{e}}$). When even ${{N}_{e}}$ is given, that ${{(\forall x{}_{i})}_{\le {{N}_{e}}-1}}$ and the characteristic function of $P(n)$ are primitive recursion function too.

Finally, the judgment result obtained must be that every even ${{N}_{e}}$ exists as follow form.
                 \[f:Mod\,\overset{\equiv }{\mathop{X}}\,(o)\to Mod\,\overset{\equiv }{\mathop{X}}\,(p)\]

That is to say, it exist $Mod\,\overset{\equiv }{\mathop{X}}\,(p)$ = (${{N}_{e}}{{(p)}^{{}}}{\;}{{\underline{{\bar{\equiv }}}}_{{\;}}}Y{{_{p}^{{}}}^{{}}}(\bmod\, X_{p}^{{}})$). So the theorem 4.2 is true.

\textbf{Theorem 4.3.} As a three operation equivalence determination model for the existence of even Goldbach conjecture, all computer recursion is solvable.

\textbf{Proof. } According to the relation model of $Mod\,\overset{\equiv }{\mathop{X}}\,(o)=({{N}_{e}}{{(o)}^{{\;}}}{{\underline{{\bar{\equiv }}}}_{{\;}}}{{Y}_{o}}^{{}}(\bmod\, {{X}_{o}}))$, by the some suitable converting, the result have:

\textbf{Basis model 1:} If ${{S}^{(1)}}=[{{N}_{e}}\equiv {{(y\underline{+}k)}^{{}}}(\bmod \,(x\bar{+}k))]$ \[(1\le(x,y,k)<N{}_{e},\;(x\overline{+}k)|({{N}_{e}}-(y\underline{+}k))\]

Then, the equivalent deciding form of Even Goldbach conjecture existence is expressed as follow:
%
%
\[{(\forall {N_e} \ge 6)_{{N_e} - is - given}}^{}{({\exists ^{}}(x\underline  +  k))_{\tiny{\begin{array}{*{20}{c}}
{\begin{array}{*{20}{c}}
{}&{}
\end{array}\left\{ {\begin{array}{*{20}{c}}
{x = 2n - 1 \le {N_e} - 1}\qquad\qquad\\
{k = 2n < {N_e}\qquad\quad\begin{array}{*{20}{c}}
{}&{}
\end{array}}
\end{array}} \right.}\\
{o{r^{}}\left\{ {\begin{array}{*{20}{c}}
{x = 2n < N{}_e\begin{array}{*{20}{c}}
{}&{}
\end{array}}\\
{k = 2n - 1 \le {N_e} - 1}\quad
\end{array}} \right.}
\end{array}}}}_{}{({\exists ^{}}(y\overline  +  k))_{\tiny{\begin{array}{*{20}{c}}
{\begin{array}{*{20}{c}}
{}&{}
\end{array}\left\{ {\begin{array}{*{20}{c}}
{y = 2n - 1 \le {N_e} - 1}\qquad\\
{k = 2n < {N_e}\quad\begin{array}{*{20}{c}}
{}&{}
\end{array}}
\end{array}} \right.}\\
{\begin{array}{*{20}{c}}\\

\end{array}{or}\left\{ {\begin{array}{*{20}{c}}
{y = 2n < {N_e}\begin{array}{*{20}{c}}
{}&{}
\end{array}}\\
{k = 2n - 1 \le {N_e} - 1}\quad
\end{array}} \right.}
\end{array}}}}\]
    \[[p(x\underline{+}k)\cap p(y\bar{+}k)\cap (((x\underline{+}k)+(y\bar{+}k))=2n+4\equiv {{N}_{e}})] (n\ge 1)\]

Where:
\[{P_{}}(n){ = ^{}}{({\exists ^{}}(x\underline  +  k))_{\tiny{\begin{array}{*{20}{c}}
{\begin{array}{*{20}{c}}
{}&{}
\end{array}\left\{ {\begin{array}{*{20}{c}}
{x = 2n - 1 \le {N_e} - 1}\qquad\qquad\\
{k = 2n < {N_e}\qquad\quad\begin{array}{*{20}{c}}
{}&{}
\end{array}}
\end{array}} \right.}\\
{or\left\{ {\begin{array}{*{20}{c}}
{x = 2n < N{}_e\begin{array}{*{20}{c}}
{}&{}
\end{array}}\\
{k = 2n - 1 \le {N_e} - 1}\quad
\end{array}} \right.}
\end{array}}}}_{}{({\exists ^{}}(y\overline  +  k))_{\tiny{\begin{array}{*{20}{c}}
{\begin{array}{*{20}{c}}
{}&{}
\end{array}\left\{ {\begin{array}{*{20}{c}}
{y = 2n - 1 \le {N_e} - 1}\qquad\qquad\\
{k = 2n < {N_e}\qquad\quad\begin{array}{*{20}{c}}
{}&{}
\end{array}}
\end{array}} \right.}\\
{\begin{array}{*{20}{c}}

\end{array}{or}\left\{ {\begin{array}{*{20}{c}}
{y = 2n < {N_e}\begin{array}{*{20}{c}}
{}&{}
\end{array}}\\
{k = 2n - 1 \le {N_e} - 1}\quad
\end{array}} \right.}
\end{array}}}}\]
\[[p(x\underline{+}k)\cap p(y\bar{+}k)\cap (((x\underline{+}k)+(y\bar{+}k))=2n+4\equiv {{N}_{e}})](n\ge 1)\]
It is the predicate relation.

\textbf{Basis model 2:}  If ${{S}^{(2)}}=[({{N}_{e}}-K){\;}{{\equiv }_{{}}}-K(\bmod \, {{N}_{e}})],(K=1,3,\cdots ,{{N}_{e}}-1)$, then the equivalent deciding form of even Goldbach conjecture existence is expressed as follow:
\[{{(\forall {{N}_{e}}\ge 6)}_{{{N}_{e}}-is-given}}{{(\exists ({{N}_{e}}-K))}_{1\le K\le {{N}_{e}}-1}}{{(\exists K)}_{1\le K\le {{N}_{e}}-1}}\]
         \[[p({{N}_{e}}-K)\cap p(K)\cap ((({{N}_{e}}-K)+K)=2n+4\equiv {{N}_{e}})]\;(n\ge 1)\]

Where:  $P(n)={{(\exists ({{N}_{e}}-K))}_{1\le K\le {{N}_{e}}-1}}{{(\exists K)}_{1\le K\le {{N}_{e}}-1}}$
                \[[p({{N}_{e}}-K)\cap p(K)\cap ((({{N}_{e}}-K)+K)=2n+4\equiv {{N}_{e}})]\;(n\ge 1)\]
It is the predicate relation.

\textbf{Basis model 3:}  If ${{S}^{(3)}}=[K{{\;}{\equiv }^{{\;}}}(K-{{N}_{e}})(\bmod\, N)],(K=1,3,\cdots ,{{N}_{e}}-1)$ then, the equivalent deciding form of even Goldbach conjecture existence is expressed as follow:
\[{(\forall {N_e} \ge 6)_{{N_e} - is - given}}{(\exists K)_{1 \le K \le {N_e} - 1}}{(\exists (K - {N_e}))_{1 \le K \le {N_e} - 1}}\]
\[[p(K) \cap p(|K - {N_e}|) \cap ((K - (K - {N_e})) = 2n + 4 \equiv {N_e})]\]

Where:
$P(n) = {(\exists K)_{1 \le K \le {N_e} - 1}}{(\exists (K - {N_e}))_{1 \le K \le {N_e} - 1}}$
\[[p(K) \cap p(|K - {N_e}|) \cap ((K - (K - {N_e})) = 2n + 4 \equiv {N_e})] {{N}_{e}})]\;(n\ge 1)\]

The description and proof of the predicate $p(n)$ may be referred the theorem 4.1 and theorem 4.2, $p(n)$ is all primary recursion predicate within the independence closed${{N}_{e}}$. Therefore, it may obtained from the last conclusion as following:

In the independent closed ${{N}_{e}}$, even Goldbach conjecture is the computer recursion solvable.  For any positive even ${{N}_{e}}\ge 6$, its solution is sure to exist odd complete congruence expressions. i.e.
                  \[Mod\,\overset{\equiv }{\mathop{X}}\,(o)=(({{N}_{e}}{{(o)}^{{\;}}}{{\underline{{\bar{\equiv }}}}_{{\;}}}{{Y}_{o}}^{{}}(\bmod\, {{X}_{o}}))\]

Or more exactly to say that, its solution is sure to exist $Mod\,\overset{\equiv }{\mathop{X}}\,(p)$ and $Mod\,\overset{\equiv }{\mathop{{{X}^{+}}}}\,(p)$, as well as $\bmod \,\overset{\equiv }{\mathop{M}}\,({{N}_{e}})$, because their operation computing are all within the independence closed ${{N}_{e}}$. (The proof end)

Above the proof show that the prime matching rule algorithm by the designed controller in model $G{{N}_{e}}TM$, within the independence closed ${{N}_{e}}$, even Goldbach conjecture is the computer recursion solvable. Any given even ${{N}_{e}}$ whether if exist the result of the prime matching, and the machine is all computable within the finite number of steps.

\section{ Halting problem not existing in the model }
Halting problem existence of the model $G{{N}_{e}}TM$ whether if existing? It directly about infinite judgment question for even Goldbach conjecture existence. If the halting problem not existing, it means that the machine not halting. At this time ${{q}_{i}}\subseteq T\;(i\ge 1)$, the machine continue keep run status. The even number result of the input at infinite tape are all true, i.e., $T/{{N}_{ei}}\;(i\ge 1)$, the primes matching pairs of every even is all existed, its result is all true. Only if the matching not halting, then the existing of even Goldbach conjecture is infinity. Otherwise, if the machine appears halting phenomenon (it can be checked). When ${{q}_{j}}\subseteq F\;(j=1)$, the result of input even number at the infinite tape is false, i.e., $F/{{N}_{ej}}\;(j\ge 1)$, it means that the even Goldbach conjecture not existence, this proposition is not established. Therefore, the infinite judgment problem of even Goldbach conjecture existence,the results also conclude that it is equivalent to the halting problem of model $G{{N}_{e}}TM$ proved.

The next part continues to discuss these detailed content, any even ${{N}_{e}}$ can be constructed equivalent proof that unique existing within the model $Mod\,\overset{\equiv }{\mathop{X}}\,(p)$ is given, it indicates that the halting problem for the model $G{{N}_{e}}TM$ does not exist.

\textbf{Definition 5.1.} Suppose that the odd positive integers of $\{X={{x}_{i}},Y={{y}_{j}}\}$ $(i,j=1,3,\cdots ,2n-1;n\ge 1)$ is the target permutation scheme element of the row and column. If the target result satisfies the ${{N}_{e}}=2n,n\ge 1$, then call the set of $S_{O}^{{}}({{N}_{e}})$ is the total combined solution matrix of two add sum relationship of the target matching below the permutation scheme of $X$ and $Y$. It is written as :

\[\begin{array}{l}
{\kern 1pt} {\kern 1pt} {\kern 1pt} {\kern 1pt} {\kern 1pt} {\kern 1pt} {\kern 1pt} {\kern 1pt} {\kern 1pt} {\kern 1pt} {\kern 1pt} {\kern 1pt} {\kern 1pt} {\kern 1pt} {\kern 1pt} {\kern 1pt} {\kern 1pt} {\kern 1pt} {\kern 1pt} {\kern 1pt} {\kern 1pt} {\kern 1pt} {\kern 1pt} {\kern 1pt} {\kern 1pt} {\kern 1pt} {\kern 1pt} {\kern 1pt} {\kern 1pt} {\kern 1pt} {\kern 1pt} {\kern 1pt} {\kern 1pt} {\kern 1pt} {\kern 1pt} {\kern 1pt} {\kern 1pt} {\kern 1pt} {\kern 1pt} {\kern 1pt} {\kern 1pt} {\kern 1pt} {\kern 1pt} {\kern 1pt} {\kern 1pt} {\kern 1pt} {\kern 1pt} {\kern 1pt} {\kern 1pt} {\kern 1pt} {\kern 1pt} {\kern 1pt} {\kern 1pt} {\kern 1pt} {\kern 1pt} {\kern 1pt} {\kern 1pt} {\kern 1pt} {\kern 1pt} {\kern 1pt} {\kern 1pt} {\kern 1pt} {\kern 1pt} {\kern 1pt} {\kern 1pt} {\kern 1pt} {\kern 1pt} {\kern 1pt} {\kern 1pt} {\kern 1pt} {\kern 1pt} {\kern 1pt} {\kern 1pt} {\kern 1pt} {\kern 1pt} {\kern 1pt} {\kern 1pt} {\kern 1pt} {\kern 1pt} {\kern 1pt} {\kern 1pt} {\kern 1pt} {\kern 1pt} {\kern 1pt} {\kern 1pt} {\kern 1pt} {\kern 1pt} {\kern 1pt} {\kern 1pt} {\kern 1pt} {\kern 1pt} {\kern 1pt} {\kern 1pt} {\kern 1pt} {\kern 1pt} {\kern 1pt} {\kern 1pt} {\kern 1pt} {\kern 1pt} {\kern 1pt} {\kern 1pt} {\kern 1pt} {\kern 1pt} {\kern 1pt} {\kern 1pt} {\kern 1pt} {\kern 1pt} {\kern 1pt} {\kern 1pt} {\kern 1pt} {\kern 1pt} {\kern 1pt} {\kern 1pt} {\kern 1pt} {\kern 1pt} {\kern 1pt} \begin{array}{*{20}{c}}
{{\kern 1pt} {\kern 1pt} {\kern 1pt} {\kern 1pt} {\kern 1pt} {\kern 1pt} {\kern 1pt} {\kern 1pt} {\kern 1pt} {\kern 1pt} {\kern 1pt} {\kern 1pt} {\kern 1pt} {y_1}}&{{\kern 1pt} {\kern 1pt} {\kern 1pt} {\kern 1pt} {\kern 1pt} {\kern 1pt} {\kern 1pt} {\kern 1pt} {\kern 1pt} {\kern 1pt} {y_3}}&{{\kern 1pt} {\kern 1pt} {\kern 1pt}  \cdots }&{{\kern 1pt} {\kern 1pt} {\kern 1pt} {\kern 1pt} {y_{2n - 3}}}&{{\kern 1pt} {\kern 1pt} {\kern 1pt} {\kern 1pt} {\kern 1pt} {\kern 1pt} {\kern 1pt} {\kern 1pt} {\kern 1pt} {\kern 1pt} {\kern 1pt} {\kern 1pt} {\kern 1pt} {y_{2n - 1}}}
\end{array}\\
{S_0}({N_e}) =  < X,Y >  = \begin{array}{*{20}{c}}
{{x_1}}\\
{{x_1}}\\
 \vdots \\
{{x_{2n - 3}}}\\
{{x_{2n - 1}}}
\end{array}\left[ {\begin{array}{*{20}{c}}
{{s_{11}}}&{{s_{13}}}& \ldots &{{s_{1(2n - 3)}}}&{{s_{1(2n - 1)}}}\\
{{s_{31}}}&{{s_{33}}}& \ldots &{{s_{3(2n - 3)}}}& \ldots \\
 \vdots & \vdots & \vdots & \vdots & \vdots \\
{{s_{(2n - 3)1}}}&{{s_{11}}}& \ldots & \ldots & \ldots \\
{{s_{(2n - 1)1}}}& \ldots & \ldots & \ldots &{{s_{(2n - 1)(2n - 1)}}}
\end{array}} \right]
\end{array}\]

\textbf{Definition 5.2. } In $S_{O}^{{}}({{N}_{e}})$, for every ${{N}_{e}}=2n,n\ge 1$, if ${{N}_{e}}={{S}_{ij}}\equiv ({{x}_{i}}+{{y}_{j}}),$ $(1\le i,j\le 2n-1)$, $\max {{S}_{ij}}\equiv ({{x}_{2n-1}}+1)$ $\equiv ({{y}_{2n-1}}+1)$, and in order to satisfy every even number below the permutation scheme of ${{X}^{{}}}an{{d}^{{}}}Y$, two add sum relationship of the target matching is all exist complete total marching solution. Then it is called the add sum matrix of the regular full permutation solution of even numbers matching, it is written as: $S_{O}^{+}({{N}_{e}})$.

There, $S_{O}^{+}({{N}_{e}})$ is just correspond with the matrix part in the clinodiagonal top-left of solid triangle form that integrity matching of ${{S}_{ij}}\;(1\le i,j\le 2n-1)$. Obviously, the matrix $S_{O}^{+}({{N}_{e}})$ has the trait as follow.

1) Constructivity, Symmetry, Uniqueness and expansible. Especially when $n\to \infty $, $S_{O}^{+}({{N}_{e}})$ is also an ideal can be constructed, that is: $S_{O}^{+\infty }({{N}_{e}}=2n)$ = $<{{X}^{+\infty }},{{Y}^{+\infty }}{{>}^{{}}}$. Because even number if the elements of $X$ and $Y$ are an infinite set, for the composition matching results can with ${{N}_{e}}$ to create the relationships of the element add sum of one by one corresponding matching. So $S_{O}^{+\infty }({{N}_{e}})$ is also the matrix of countable infinite even number adds sum.

2) Those even numbers lie in each deputy diagonal of the matrix $S_{o}^{+}({{N}_{e}})$, it pledges even is the same, and different each other, and they forms continuous rank for $2n\;(n\ge 1)$ sequence within given even numbers.

3) In the matrix $S_{o}^{+}({{N}_{e}})$, it each number that even of continue rank lie in every deputy diagonal, which just correspond exactly to natural number sequence: 1,2,3,4,5,
$\cdots ,n$, $n\ge 1$.

\textbf{Definition 5.3.}  In the add sum relationships of the target matching of $S_{O}^{+}({{N}_{e}})$, if only select the corresponding to complete matching solutions result of all add sum relations, and it is the permutation of max even ${{N}_{e}}$, then call the max even ${{N}_{e}}$ is the matrix of the even add sum for maximum regular complete full rank. It is written as:
%
%
\[\begin{array}{l}
{\kern 1pt} {\kern 1pt} {\kern 1pt} {\kern 1pt} {\kern 1pt} {\kern 1pt} {\kern 1pt} {\kern 1pt} {\kern 1pt} {\kern 1pt} {\kern 1pt} {\kern 1pt} {\kern 1pt} {\kern 1pt} {\kern 1pt} {\kern 1pt} {\kern 1pt} {\kern 1pt} {\kern 1pt} {\kern 1pt} {\kern 1pt} {\kern 1pt} {\kern 1pt} {\kern 1pt} {\kern 1pt} {\kern 1pt} {\kern 1pt} {\kern 1pt} {\kern 1pt} {\kern 1pt} {\kern 1pt} {\kern 1pt} {\kern 1pt} {\kern 1pt} {\kern 1pt} {\kern 1pt} {\kern 1pt} {\kern 1pt} {\kern 1pt} {\kern 1pt} {\kern 1pt} {\kern 1pt} {\kern 1pt} {\kern 1pt} {\kern 1pt} {\kern 1pt} {\kern 1pt} {\kern 1pt} {\kern 1pt} {\kern 1pt} {\kern 1pt} {\kern 1pt} {\kern 1pt} {\kern 1pt} {\kern 1pt} {\kern 1pt} {\kern 1pt} {\kern 1pt} {\kern 1pt} {\kern 1pt} {\kern 1pt} {\kern 1pt} {\kern 1pt} {\kern 1pt} {\kern 1pt} {\kern 1pt} {\kern 1pt} {\kern 1pt} {\kern 1pt} {\kern 1pt} {\kern 1pt} {\kern 1pt} {\kern 1pt} {\kern 1pt} {\kern 1pt} {\kern 1pt} {\kern 1pt} {\kern 1pt} {\kern 1pt} {\kern 1pt} {\kern 1pt} {\kern 1pt} {\kern 1pt} {\kern 1pt} {\kern 1pt} {\kern 1pt} {\kern 1pt} {\kern 1pt} {\kern 1pt} {\kern 1pt} {\kern 1pt} {\kern 1pt} {\kern 1pt} {\kern 1pt} {\kern 1pt} {\kern 1pt} {\kern 1pt} {\kern 1pt} {\kern 1pt} {\kern 1pt} {\kern 1pt} {\kern 1pt} {\kern 1pt} {\kern 1pt} {\kern 1pt} {\kern 1pt} {\kern 1pt} {\kern 1pt} {\kern 1pt} {\kern 1pt} {\kern 1pt} {\kern 1pt} {\kern 1pt} {\kern 1pt} {\kern 1pt} {\kern 1pt} \begin{array}{*{20}{c}}
{{\kern 1pt} {\kern 1pt} {\kern 1pt} {\kern 1pt} {\kern 1pt} {\kern 1pt} {\kern 1pt} {\kern 1pt} {\kern 1pt} {\kern 1pt} {\kern 1pt} {\kern 1pt} {\kern 1pt} {\kern 1pt} {\kern 1pt} {\kern 1pt} {\kern 1pt} {\kern 1pt} {\kern 1pt} {\kern 1pt} {\kern 1pt} {\kern 1pt} {\kern 1pt} {\kern 1pt} {\kern 1pt} {\kern 1pt} {\kern 1pt} {\kern 1pt} {\kern 1pt} {\kern 1pt} {\kern 1pt} {\kern 1pt} {\kern 1pt} {y_1}}&{{\kern 1pt} {\kern 1pt} {\kern 1pt} {\kern 1pt} {\kern 1pt} {\kern 1pt} {\kern 1pt} {\kern 1pt} {\kern 1pt} {\kern 1pt} {\kern 1pt} {\kern 1pt} {\kern 1pt} {\kern 1pt} {\kern 1pt} {\kern 1pt} {\kern 1pt} {y_3}}&{{\kern 1pt} {\kern 1pt} {\kern 1pt} {\kern 1pt} {\kern 1pt} {\kern 1pt} {\kern 1pt} {\kern 1pt}  \cdots }&{{\kern 1pt} {\kern 1pt} {\kern 1pt} {\kern 1pt} {\kern 1pt} {y_{2n - 3}}}&{{\kern 1pt} {\kern 1pt} {\kern 1pt} {\kern 1pt} {\kern 1pt} {\kern 1pt} {y_{2n - 1}}}
\end{array}\\
S_0^ + (\max {N_e}) =  < X,Y >  = \begin{array}{*{20}{c}}
{{x_1}}\\
{{x_1}}\\
 \vdots \\
{{x_{2n - 3}}}\\
{{x_{2n - 1}}}
\end{array}\left[ {\begin{array}{*{20}{c}}
0&0& \ldots &{{s_{1(2n - 3)}}}&{{s_{1(2n - 1)}}}\\
0&0& \ldots &{{s_{3(2n - 3)}}}&{}\\
 \vdots & \vdots & \vdots &{}&{}\\
0&{{s_{(2n - 3)3}}}&{}&{}&{}\\
{{s_{(2n - 1)1}}}&{}&{}&{}&{}
\end{array}} \right]
\end{array}\]

Actually, the matrix $S_{O}^{+}(\max {{N}_{e}})$ is also show that the result of two odd integer add sum at upper clinodiagonal, i.e,. it is formed the part that the combination result of maximum complete even numbers. So there is:
\[S_{O}^{+}(\max {{N}_{e}})\in S_{o}^{+}({{N}_{e}})\in {{S}_{0}}({{N}_{e}}),\]
\[S_{O}^{+}(\max {{N}_{e}})\sim Mod\,\overset{\equiv }{\mathop{X}}\,(o).\]

\textbf{Lemma 5.1.} Every even number ${{N}_{e}}$ can all independently construct the matrix of $S_{O}^{+}(\max {{N}_{e}})$.

\textbf{Proof.} In $S_{O}^{{}}({{N}_{e}})$, if only to be determined $\max {{S}_{ij}}\equiv ({{x}_{2n-1}}+1)$ $\equiv ({{y}_{2n-1}}+1)$ is given,   every determined ${{N}_{e}}$  there is all an independent $S_{O}^{+}(\max {{N}_{e}})$ corresponding with it, and satisfies the even number sequence {2 $\subset $ 4 $\subset $ 6 $\subset \cdots $ $\subset $ $S_{O}^{+}(\max {{N}_{e}}-2)$ $\subset $ $S_{O}^{+}(\max {{N}_{e}})$}. On the contrary, every $S_{O}^{+}(\max {{N}_{e}})$ sure cover all the even numbers that less than it. Because of $S_{O}^{+}(\max {{N}_{e}})$ $\in S_{o}^{+}({{N}_{e}})$. $S_{O}^{+}(\max {{N}_{e}})$ is an individual expression, and $S_{o}^{+}({{N}_{e}})$ is the total expression within given ${{N}_{e}}$.

\textbf{Lemma 5.2.} Suppose that $\sum\nolimits_{\varphi }{({{N}_{e}})}$ is every even ${{N}_{e}}$ in $S_{o}^{+}({{N}_{e}})$ exist the full permutation solutions number of the even add sum relationships for has the complete matching , then there is $\sum\nolimits_{\varphi }{({{N}_{e}})}$ = ${{N}_{e}}/2$, and it is equivalent to the deduction 2.1.

\textbf{Proof. }  According to the definition 5.1, 5.2 and 5.3 known that, in the matrix $S_{o}^{+}({{N}_{e}})$, every even ${{N}_{e}}$ exist all full permutation matching solution of the regular even add sum relations.  Once ${{N}_{e}}$ is given, ${{s}_{i,j}}\;(1\le i,j\le (2n-1)$ of all the corresponding add sum relationships are also determined. And there is the relation existing of even add sum for $\{{{s}_{(2n-1)1}},{{s}_{(2n-3)3}},\cdots ,$ ${{s}_{3(2n-3)}},{{s}_{1(2n-1)}}\}$ full permutations solution of complete matching, that is $\sum\nolimits_{\varphi }{({{S}_{ij}})}=$ $\sum\nolimits_{\varphi }{({{x}_{i}},{{y}_{j}})}=(2n-i)+j$, $(1\le i,j\le 2n-1)$, when $i=j$, then have $\sum\nolimits_{\varphi }{({{S}_{ij}}})=2n$. Because the in matching composition of ${{s}_{ij}}=({{x}_{i}},{{y}_{j}})$, the element $i,j$ only select odd number, therefore there is $\sum\nolimits_{\varphi }{({{S}_{ij}})}=2n/2={{N}_{e}}/2$.
%
%

Another according to the definition 2.3, the lemma 2.2, which can be known with the deduction 2.1 is equivalent.

\textbf{Deduction 5.1.} If variable $x$ and $y$ which can from the smallest combination $({{x}_{3}},{{y}_{3}})$ starting in the matrix $S_{o}^{+}({{N}_{e}})$, and every even ${{N}_{e}}$ exist total numbers for full arrangement result of even add sum relation of complete pairing, then have $\sum\nolimits_{\varphi }^{*}{({{N}_{e}})}=[({{N}_{e}}/2)-2]$.

\textbf{Proof. } Because of when the variable $x$ and $y$ which can from the smallest combination $({{x}_{3}},{{y}_{3}})$ starting, it correspond matching solution have $({{x}_{3}},{{y}_{3}})=6$, and even number 6 lie in the deputy diagonal of the matrix, it correspond number just from number 1 begin. Obvious, given the matrix matching solution missing $({{x}_{1}},{{y}_{1}})=2$ and $({{x}_{1}},{{y}_{3}})=({{x}_{3}},{{y}_{1}})=4$, thus result only have $\sum\nolimits_{\varphi }^{*}{({{N}_{e}})}=[({{N}_{e}}/2)-2]$ can satisfying the requirement that the numbers exist the  sequence for $1,2,3,\cdots ,k$ $(1\le k\le ({{{N}_{e}}}/{2}\;)-2)$ at clinodiagonal.

\textbf{Definition 5.4.} Let ${{N}_{e}}(p)=\{{{p}_{i}}(x),{{p}_{j}}(y)|:(i,j=1,2,\cdots ,r)\}$(the element $x,y$ is  respectively correspond to the row and column vectors) is the odd primes set that not more than given even number${{N}_{e}}$. If $S_{P}^{{}}({{N}_{e}})=$ $<{{p}_{i}}(x),{{p}_{j}}(y)>$ is the non-empty set of the add sum relationships of prime matching, then call ${{S}_{P}}({{N}_{e}})$ is the matrix of prime add summation. It is written as:
\[\begin{array}{l}
{\kern 1pt} {\kern 1pt} {\kern 1pt} {\kern 1pt} {\kern 1pt} {\kern 1pt} {\kern 1pt} {\kern 1pt} {\kern 1pt} {\kern 1pt} {\kern 1pt} {\kern 1pt} {\kern 1pt} {\kern 1pt} {\kern 1pt} {\kern 1pt} {\kern 1pt} {\kern 1pt} {\kern 1pt} {\kern 1pt} {\kern 1pt} {\kern 1pt} {\kern 1pt} {\kern 1pt} {\kern 1pt} {\kern 1pt} {\kern 1pt} {\kern 1pt} {\kern 1pt} {\kern 1pt} {\kern 1pt} {\kern 1pt} {\kern 1pt} {\kern 1pt} {\kern 1pt} {\kern 1pt} {\kern 1pt} {\kern 1pt} {\kern 1pt} {\kern 1pt} {\kern 1pt} {\kern 1pt} {\kern 1pt} {\kern 1pt} {\kern 1pt} {\kern 1pt} {\kern 1pt} {\kern 1pt} {\kern 1pt} {\kern 1pt} {\kern 1pt} {\kern 1pt} {\kern 1pt} {\kern 1pt} {\kern 1pt} {\kern 1pt} {\kern 1pt} {\kern 1pt} {\kern 1pt} {\kern 1pt} {\kern 1pt} {\kern 1pt} {\kern 1pt} {\kern 1pt} {\kern 1pt} {\kern 1pt} {\kern 1pt} {\kern 1pt} {\kern 1pt} {\kern 1pt} {\kern 1pt} {\kern 1pt} {\kern 1pt} {\kern 1pt} {\kern 1pt} {\kern 1pt} {\kern 1pt} {\kern 1pt} {\kern 1pt} {\kern 1pt} {\kern 1pt} {\kern 1pt} {\kern 1pt} {\kern 1pt} {\kern 1pt} {\kern 1pt} {\kern 1pt} {\kern 1pt} {\kern 1pt} {\kern 1pt} {\kern 1pt} {\kern 1pt} {\kern 1pt} {\kern 1pt} {\kern 1pt} {\kern 1pt} {\kern 1pt} {\kern 1pt} {\kern 1pt} {\kern 1pt} {\kern 1pt} {\kern 1pt} {\kern 1pt} {\kern 1pt} {\kern 1pt} {\kern 1pt} {\kern 1pt} {\kern 1pt} {\kern 1pt} {\kern 1pt} {\kern 1pt} {\kern 1pt} {\kern 1pt} {\kern 1pt} {\kern 1pt} {\kern 1pt} {\kern 1pt} {\kern 1pt} {\kern 1pt} {\kern 1pt} {\kern 1pt} {\kern 1pt} {\kern 1pt} {\kern 1pt} {\kern 1pt} {\kern 1pt} {\kern 1pt} {\kern 1pt} {\kern 1pt} {\kern 1pt} {\kern 1pt} {\kern 1pt} {\kern 1pt} {\kern 1pt} {\kern 1pt} {\kern 1pt} {\kern 1pt} {\kern 1pt} {\kern 1pt} {\kern 1pt} {\kern 1pt} {\kern 1pt} {\kern 1pt} {\kern 1pt} {\kern 1pt} {\kern 1pt} {\kern 1pt} {\kern 1pt} {\kern 1pt} {\kern 1pt} {\kern 1pt} {\kern 1pt} {\kern 1pt} {\kern 1pt} {\kern 1pt} {\kern 1pt} {\kern 1pt} {\kern 1pt} {\kern 1pt} {\kern 1pt} \begin{array}{*{20}{c}}
{{p_1}(y)}&{{p_2}(y)}& \cdots &{{\kern 1pt} {\kern 1pt} {\kern 1pt} {\kern 1pt} {p_r}(y)}
\end{array}\\
{S_p}({N_e}) ={\,} < {p_i}(x),{p_j}(y) >{\,}  = \begin{array}{*{20}{c}}
{{p_1}(x)}\\
{{p_2}(x)}\\
 \vdots \\
{{p_r}(x)}
\end{array}{\kern 1pt} {\kern 1pt} {\kern 1pt} {\kern 1pt} \left[ {{\kern 1pt} {\kern 1pt} \begin{array}{*{20}{c}}
{{\kern 1pt} {\kern 1pt} {\kern 1pt} {\kern 1pt} {\kern 1pt} {\kern 1pt} s_{11}^{'}{\kern 1pt} {\kern 1pt} {\kern 1pt} {\kern 1pt} {\kern 1pt} {\kern 1pt} {\kern 1pt} {\kern 1pt} {\kern 1pt} {\kern 1pt} {\kern 1pt} {\kern 1pt} {\kern 1pt} {\kern 1pt} {\kern 1pt} {\kern 1pt} }&{s_{12}^{'}}&{{\kern 1pt} {\kern 1pt} {\kern 1pt} {\kern 1pt} {\kern 1pt} {\kern 1pt}  \cdots }&{{\kern 1pt} {\kern 1pt} {\kern 1pt} {\kern 1pt} {\kern 1pt} s_{1r}^{'}{\kern 1pt} {\kern 1pt} {\kern 1pt} {\kern 1pt} {\kern 1pt} {\kern 1pt} {\kern 1pt} {\kern 1pt} {\kern 1pt} {\kern 1pt} }\\
{{\kern 1pt} {\kern 1pt} {\kern 1pt} {\kern 1pt} {\kern 1pt} {\kern 1pt} s_{21}^{'}{\kern 1pt} {\kern 1pt} {\kern 1pt} {\kern 1pt} {\kern 1pt} {\kern 1pt} {\kern 1pt} {\kern 1pt} {\kern 1pt} {\kern 1pt} {\kern 1pt} {\kern 1pt} {\kern 1pt} {\kern 1pt} {\kern 1pt} {\kern 1pt} }&{s_{22}^{'}}&{{\kern 1pt} {\kern 1pt} {\kern 1pt} {\kern 1pt} {\kern 1pt} {\kern 1pt}  \cdots }&{{\kern 1pt} {\kern 1pt} {\kern 1pt} {\kern 1pt} {\kern 1pt} s_{2r}^{'}{\kern 1pt} {\kern 1pt} {\kern 1pt} {\kern 1pt} {\kern 1pt} {\kern 1pt} {\kern 1pt} {\kern 1pt} {\kern 1pt} {\kern 1pt} }\\
 \vdots & \vdots &{{\kern 1pt} {\kern 1pt} {\kern 1pt} {\kern 1pt} {\kern 1pt}  \vdots }& \vdots \\
{s_{r1}^{'}{\kern 1pt} {\kern 1pt} {\kern 1pt} {\kern 1pt} {\kern 1pt} {\kern 1pt} {\kern 1pt} }&{s_{r1}^{'}}&{{\kern 1pt} {\kern 1pt} {\kern 1pt} {\kern 1pt}  \cdots }&{s_{rr}^{'}{\kern 1pt} {\kern 1pt} {\kern 1pt} {\kern 1pt} {\kern 1pt} }
\end{array}} \right]
\end{array}\]

Obviously, the existence of the matrix ${{S}_{P}}({{N}_{e}})$ is also unique, and it is ideal constructive. The element $s_{ij}^{'}$ is distributed symmetrically about the center of the main diagonal. Because even if the element ${{p}_{i}}{{(x)}^{{}}}$ and ${{}^{{}}}{{p}_{j}}(y)$ are infinite sets, for the reason that this result can with ${{N}_{e}}$ build one by one the relationships of corresponding matching add sum, so $S_{p}^{{}}({{N}_{e}})$ is also a kind of relationship matrix of countable infinite add sum result.

\textbf{Axiom 5.1.} The existence of prime number is infinite, it can be described as: ${{p}^{\infty }}={{\{2={{p}_{1}}{{,}^{{}}}3={{p}_{2}}{{,}^{{}}}5={{p}_{3}},\cdots ,{{p}_{r}}{{,}^{{}}}{{p}_{r+1}},\cdots \}}^{\infty }}$.

\textbf{Deduction 5.2.} If ${{p}^{+\infty }}$ is the set of countable infinite primes, then have the even matrix of prime add sum that to consist of $S_{P}^{+\infty }({{N}_{e}})=$ $<p_{i}^{+\infty }(x),p_{j}^{+\infty }(y)>$ is countable infinitely extensible and constructive.

\textbf{Proof. } It's known that ${{p}^{\infty }}$ is an infinite set from Axiom 5.1. Because the matching of $<p_{i}^{+\infty }(x),p_{j}^{+\infty }(y)>$ can with $S_{P}^{+\infty }$ build the relationships by one to one corresponding element add sum, so $S_{P}^{+\infty }$ is the relationship matrix of the element add sum in countable infinite set. Of course, it's also countable infinitely extensible and constructive. so the deduction 5.2 is established.

\textbf{Lemma 5.3.} The matrix ${{S}_{P}}({{N}_{e}})$ is the implicit matrix of the matrix $S_{O}^{{}}({{N}_{e}})$.

\textbf{Proof. } According to the definition 5.1, If ${{N}_{e}}=2{{n}^{{}}}(n\ge 1)$ is given, then have all odd prime existing in ${{N}_{e}}$, i.e., ${{N}_{e}}(p)=$ $\{3={{p}_{1}}<{{p}_{2}}<\cdots <{{p}_{i}}<\cdots <{{p}_{r}}\le {{N}_{e}}-1\; =\; 2n-1\}$ $(1\le i\le r)$. And the correspond to the row and column vectors of $\{X,Y\}$ in the matrix $S_{O}^{{}}({{N}_{e}})$, which respectively is always the subset of odd integers :${{N}_{e}}(X)={{N}_{e}}(Y)$ = $\{1,3,\cdots ,2n-1\}$, it situated the even relation sequence of the matching add sums of upper main diagonal are complete. And ${{N}_{e}}(p) =$ ${{p}_{i}}(1\le i\le r)\in $ $\{X,Y\}$, its even relation sequence of the matching result is incomplete. Additionally, let ${{N}_{e}}(d\overline{p})$ is represent the  non-primes set of  par element in the sets ${{N}_{e}}{{(X)}^{{}}}$ and ${{}^{{}}}{{N}_{e}}(Y)$, then have $\{$[${{N}_{e}}(X) = {{N}_{e}}(Y)$]- ${{N}_{e}}(d\overline{p})$$\}$ = ${{N}_{e}}(p)$. On the contrary,${{N}_{e}}(p)\bigcup {{N}_{e}}(d\overline{p}) = {{N}_{e}}(X) = {{N}_{e}}(Y)$. The result have ${{S}_{d}}({{N}_{e}}) = \;<{{x}_{d}},{{y}_{d}}>$ = $<{{N}_{e}}(X)-{{N}_{e}}(p){{,}_{{}}}$ ${{N}_{e}}(Y)-{{N}_{e}}(p)>$ $\Rightarrow $ $S_{O}^{{}}({{N}_{e}})-{{S}_{d}}({{N}_{e}})$ = ${{S}_{P}}({{N}_{e}})$. It shows that as long as the non-prime element relationship is removed from the matrix $S_{O}^{{}}({{N}_{e}})$, the row and column vectors corresponding to $\{X, Y\}$ are constructed, and all matching addition sums are constructed,
the result is still the matrix ${{S}_{P}}({{N}_{e}})$. Conversely, only if in the matrix ${{S}_{P}}({{N}_{e}})$, the row and column vectors of ${{p}_{i}}\;(1\le i\le r)$ increases to be deleted the element and the result of matching add sum relationships of corresponding to original row and column,  then ${{S}_{P}}({{N}_{e}})$ is restored to the matrix $S_{O}^{{}}({{N}_{e}})$, that is ${{S}_{P}}({{N}_{e}})\bigcup {{S}_{d}}({{N}_{e}})$ = $S_{O}^{{}}({{N}_{e}})$. So the lemma is established.

For example, when ${{N}_{e}}=20$, the constructed transformation  process of $S_{O}^{'}({{N}_{e}})$ $\Rightarrow $ ${{S}_{P}}({{N}_{e}})$ as shown in Fig 3.

%
%
\begin{figure}[!htb]
   \centering
   \includegraphics[width=85mm]{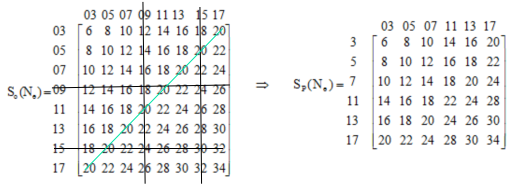}
    \caption{the conversion process example for $S_{O}^{'}({{N}_{e}})$ $\Rightarrow $ ${{S}_{P}}({{N}_{e}})$}
    \label{Machine}
\end{figure}

\textbf{Lemma 5.4. }  If $p{}_{s}$ is the maximum prime (except for $({{N}_{e}}-1)$ is prime) within given even ${{N}_{e}}$, then every even number $N_{e}^{'}\le {{N}_{e}}-2$ in the matrix ${{S}_{P}}({{N}_{e}})$, each even-matched addition is equivalent of the relationship to the model $Mod\,\overset{\equiv }{\mathop{X}}\,(p)$.

\textbf{Proof. } Suppose that $p$ is a prime , within given even ${{N}_{e}}$, ${{p}_{s}}=\max p<{{N}_{e}}-1\in {{N}_{e}}$. According the definition 5.4 has known $S{}_{p}({{N}_{e}})=\{{{p}_{i}}(x),{{p}_{j}}(y)|:(i,j=1,2,\cdots ,s)\}$\;($x,y$ corresponding the row and the column vectors ) is non-empty set matrix for the adding sum relationships of odd prime matching. Once ${{N}_{e}}$ is given, and ${{p}_{s}}$ is also given, and both of them satisfy $\max {{N}_{e}}={{p}_{s}}+q$ $(q=(3={{p}_{1}},5={{p}_{2}},\cdots ,{{p}_{g}}))$ $(g\ge 1)$ of respective adding sum relations of the complete prime matching. According to the structural characteristics of ${{S}_{P}}({{N}_{e}})$, then there are full permutations about every even exist matching adding sum within $6\le {{N}_{e}}\le $ $\max {{N}_{e}}={{P}_{S}}+q$ $(q=(3={{p}_{1}},5={{p}_{2}},\cdots ,{{p}_{g}}))$ $(g\ge 1)$. The full permutation set of this adding sum relationship of each even is just equivalent to the set of the congruence expression  $Mod\,\overset{\equiv }{\mathop{X}}\,(p)$ = [${{N}_{e}}{{(p)}^{{}}}{\;}{{\underline{{\bar{\equiv }}}}_{{\;}}}Y{{_{p}^{{}}}^{{}}}(\bmod\, X_{p}^{{}})$] in the definition 2.4, because here the element $x,y$ are a group of determined mod x permutations result of full congruence adding sum relation consisted of all the odd prime numbers of $(p{{'}_{1}},p{{'}_{2}},p{{'}_{3}},\ldots ,p{{'}_{s-2}},p{{'}_{s-1}},p{{'}_{s}})$ $(1\le s<r)$.

\textbf{Lemma 5.5.} If ${{N}_{e}}$ is given in $Mod\overset{\equiv }{\mathop{X}}\,(o)$, then there is the numbers of prime that not more than the given positive even ${{N}_{e}}$ have $\pi ({{N}_{e}})\approx {{{N}_{e}}}/{\ln {{N}_{e}}}\;$.

\textbf{Proof.}  By the prime number theorem known that $\underset{x\to \infty }{\mathop{lin}}\,(\pi (x)/(x/\ln x)=1$,let $x={{N}_{e}}$, The lemma is easy to prove .

\textbf{Deduction 5.3.} The number of prime that not more than the positive even of ${Ne}/{2}\;$, have $\pi ({{N}_{e}}/2)\approx ({{{N}_{e}}/2)}/{\ln ({{N}_{e}}/2)}\;$.

\textbf{Deduction 5.4.} In the range $[({{N}_{e}}/2)+1,{{N}_{e}}]$, there are $\pi ({{N}_{e}}-{{N}_{e}}/2)=\pi ({{N}_{e}})-\pi ({{N}_{e}}/2)$ primes existing.

\textbf{Theorem 5.1.} In the matrix ${{S}_{P}}({{N}_{e}})$, as long as there is an even ${{N}_{e}}$ not existing, then the matrix $S_{O}^{+}(\max {{N}_{e}})$ not existing, the matrix $S_{O}^{{}}({{N}_{e}})$ is also not existing.

\textbf{Proof. } Firstly, we examine the proof in $S_{O}^{+}(\max {{N}_{e}})$, if the even ${{N}_{e}}$ not found in the matrix ${{S}_{P}}({{N}_{e}})$, if and only if full permutation of the even of $\sum\nolimits_{\varphi }{({{N}_{e}})}$ numbers in $S_{O}^{+}(\max {{N}_{e}})$ is not exist.

\textbf{Necessity.} According to the lemma 5.1, every ${{N}_{e}}$ can all be uniquely constructed a corresponding matrix $S_{O}^{+}(\max {{N}_{e}})$, and have full permutation existence of the even of $\sum\nolimits_{\varphi }{({{N}_{e}})}$ = ${{{N}_{e}}}/{2}\;$ numbers.

\textbf{Sufficiency.} According to the lemma 5.3, the matrix ${{S}_{P}}({{N}_{e}})$ is the implicit matrix of the matrix $S_{O}^{{}}({{N}_{e}})$. If any even ${{N}_{e}}$ not found in the matrix ${{S}_{P}}({{N}_{e}})$, then, only when ${{N}_{e}}$ corresponding the matrix $S_{O}^{+}(\max {{N}_{e}})$ is not exist. If non-existence is true, then there are only a few possibilities.

\textbf{Situation 1.} Appoint firstly: If any even ${{N}_{e}}$ = $<{{x}_{i}},{{y}_{j}}>$$(1\le i,j\le 2n-1)$ not exist, then the full even of $\sum\nolimits_{\varphi }{({{N}_{e}})}$ numbers of adding sum relation that the corresponding row and column matching which should be all deleted. Assuming in $S_{O}^{+}(\max {{N}_{e}})$ = $\max <x,y>$, the elements of all variables $x,y$ are odd integers, that is, when $x=y=2n-1{{,}^{{}}}n\ge 1$, the deleted variable can be expressed as:

$delete\{x\}=\{d{{x}_{1}},d{{x}_{3}},d{{x}_{5}},\cdots ,d{{x}_{2n-3}},d{{x}_{2n-1}}\}$ $\Rightarrow delete{{N}_{e}}$, it is expressed shortly as: $d\{{{x}_{1\tilde{\ }2n-1}}\Rightarrow \times {{N}_{e}}\}$.

The result shows that when the variable of $x$ from ${{x}_{1}}\to {{x}_{2n-1}}$ are all odd integers, all determined even of $\sum\nolimits_{\varphi }{({{N}_{e}})}$ numbers that all corresponding to the row which should be completely deleted. Because whether the variable $y$ is prime or odd, it consisted relation matching solution ${{N}_{e}}$ = $<{{x}_{i}},{{y}_{j}}>$ of adding summation which can't satisfy the requirements, let alone all $y$ has been assumed as odd integers.
%
%

Or:$delete\{y\}=\{d{{y}_{1}},d{{y}_{3}},d{{y}_{5}},\cdots ,d{{y}_{2n-3}},d{{y}_{2n-1}}\}$ $\Rightarrow delete{{N}_{e}}$, it is expressed shortly as: $d\{{{y}_{1\tilde{\ }2n-1}}\Rightarrow \times {{N}_{e}}\}$.

The result shows that when the variable of $y$ from ${{y}_{1}}\to {{y}_{2n-1}}$ are all odd integers, the determined even of $\sum\nolimits_{\varphi }{({{N}_{e}})}$ numbers that all corresponding to the column which should be completely deleted. Because all variables $x$ of with $y$ matching are also all odd numbers, it has not one of them can satisfy the requirement.

$delete<x,y>$ express full even of $\sum\nolimits_{\varphi }{({{N}_{e}})}$ numbers that all matching adding sum relation should be completely deleted. $d\{{{y}_{1\tilde{\ }2n-1}}\Rightarrow \times {{N}_{e}}\}$ means that have the even of $\sum\nolimits_{\varphi }{({{N}_{e}})}$ = ${{{N}_{e}}}/{2}\;$ numbers in $S_{O}^{+}(\max {{N}_{e}})$ = $\max <x,y>$ which should be completely deleted, only this way to can demonstrate all $\sum\nolimits_{\varphi }{({{N}_{e}})}$ even in $S_{O}^{+}(\max {{N}_{e}})$ is not existence. If this deletion result makes not existence is true, then it makes ${{N}_{e}}$ not existence in the matrix ${{S}_{P}}({{N}_{e}})$ is possible. Obviously, this result is the contradiction with the lemma 5.3.

\textbf{Situation 2.} Assume that any even ${{N}_{e}}$ not existing $S_{O}^{+}(\max {{N}_{e}})$ = $\max <x,y>$, Only when variables of $x,y$ are all positive odd numbers which respectively in the interval of $[1,{{N}_{e}}/2-1]$\;(or\;[$1,{{{N}_{e}}}/{2}\;$]), the remnant is the positive integers (it including prime and odd numbers), thus constructed even number belongs to be deleted the target, as shown in Figure 4(a); or assume that the variables of $x,y$ are all positive odd numbers in the interval of $[({{N}_{e}}/2)+1,{{N}_{e}}-1]$\;(or[$({{{N}_{e}}}/{2}\;)+2,{{N}_{e}}-1$]), the remnant is all positive integers (it including prime and odd numbers), which corresponding range even is also deleted, as shown in Figure 4(b).
\begin{figure}
  \centering
  \includegraphics[width=.8\textwidth]{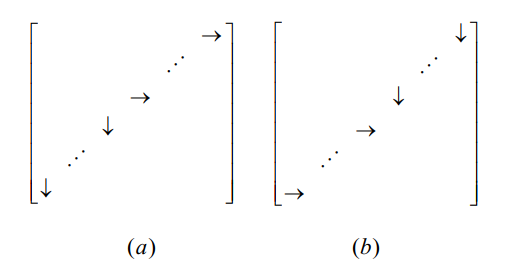} 
  \caption{The sketch map of to delete different directions} 
  \label{img} 
\end{figure}

And because in the matrix $S_{O}^{{}}({{N}_{e}})$, the distribution of even number consist of two part
${{N}_{e}}=\{4n\cup (4n-2)\}\;(n\ge 1)$. For example, $N_{e}^{1}=(4n-2)\;(n\ge 1)=\{2,6,10,14,\cdots \}$, $N_{e}^{2}=4{{n}_{{}}}\;(n\ge 1)=\{4,8,12,16,\cdots \}$, and ${{N}_{e}}=\{N_{e}^{1}\cup N_{e}^{2}\}=\{2,4,6,8,10,12,14,16,\cdots \}$. So there is odd and even the existence result of two matrix different constructions, therefore, the distribution interval will also occur corresponding change.

When the number of x and y variable is odd number, it is called odd variable matrix. About the interval distribution of x and y two part is $[1,{{{N}_{e}}}/{2}\;]$ and $[({{{N}_{e}}}/{2}\;)+2,{{N}_{e}}-1]$, or:$[1,({{{N}_{e}}}/{2}\;)-2]$ and $[({{{N}_{e}}}/{2}\;),{{N}_{e}}-1]$. When the number of x and y variable is even number, it is called even variable matrix. About the interval distribution of x and y two part is $[1,{({{N}_{e}}}/{2)-1}\;]$ and $[({{{N}_{e}}}/{2}\;)+1,{{N}_{e}}-1]$.

Therefore, according to delete model for as shown in Figure 4(a) and (b), the result is deleted severally case as follows.

1.The even variable matrix case: deleting corresponding even space ${{N}_{e}}=$$\{4n,n\ge 1\}$

(1) In the interval of $[1,({{N}_{e}}/2)-1]$, when $x,y$ are all positive odd numbers, along the direction of variable x,y delete corresponding even ${{N}_{e}}$ as shown in Figure 4(a), it has two kinds of matching add sun relation.

 $a$) X direction:

 Deleting $\{{{x}_{1\sim{\ }({{N}_{e}}/2)-1}},y{}_{({{N}_{e}}/2)+1\sim{\ }({{N}_{e}}-1)}\}$$|:(({{x}_{1}},{{y}_{{{N}_{e}}-1}}),$ $\cdots ,({{x}_{({{N}_{e}}/2)-1}},{{y}_{({{N}_{e}}/2)+1}}))$

 $b$) Y direction:

 Deleting $\{{{y}_{1\sim{\ }({{N}_{e}}/2)-1}},x{}_{({{N}_{e}}-1)\sim{\ }({{N}_{e}}/2)+1}\}$$|:(({{y}_{1}},{{x}_{{{N}_{e}}-1}}),$ $\cdots ,({{y}_{({{N}_{e}}/2)-1}},{{x}_{({{N}_{e}}/2)+1}}))$

 (2) In the interval of $[({{{N}_{e}}}/{2}\;)+1,{{N}_{e}}-1]$, when $x,y$ are all positive odd numbers, along the direction of variable x,y delete corresponding even ${{N}_{e}}$ as shown in Figure 4(b), it has also two kind of matching add sun relation.

 $a$) X direction:

 Deleting $\{{{x}_{({{N}_{e}}/2)+1\sim{\ }({{N}_{e}}-1)}},y{}_{({{N}_{e}}/2)-1\sim{\ }1)}\}$$|:(({{x}_{(N/2)+1}},{{y}_{({{N}_{e}}/2)-1}}),$$\cdots ,({{x}_{{{N}_{e}}-1}},{{y}_{1}}))$

 $b$) Y direction:

 Deleting $\{y{}_{({{N}_{e}}/2)+1\sim{\ }({{N}_{e}}-1)},{{x}_{({{N}_{e}}/2)-1\sim{\ }1}}\}$$|:(({{y}_{(N/2)+1}},{{x}_{({{N}_{e}}/2)-1}}),$$\cdots ,({{y}_{{{N}_{e}}-1}},{{x}_{1}}))$

2.The odd variable matrix case: deleting corresponding even space ${{N}_{e}}=$ $\{(4n-2),n\ge 1\}$

(1) In the interval of $[1,({{N}_{e}}/2)]$, (or: $[1,({{N}_{e}}/2)-2]$, this case is not discussion), when $x,y$ are all positive odd numbers, along the direction of variable x,y delete corresponding even ${{N}_{e}}$ as shown in Figure 4(a), it has two kind of matching add sun relation.

 $a$) X direction:

 Deleting $\{{{x}_{1\sim{\ }({{N}_{e}}/2)}},y{}_{({{N}_{e}}-1)\sim{\ }({{N}_{e}}/2)}\}$$|:(({{x}_{1}},{{y}_{{{N}_{e}}-1}}),$ $\cdots ,({{x}_{({{N}_{e}}/2)}},{{y}_{({{N}_{e}}/2)}}))$

 $b$) Y direction:

 Deleting $\{{{y}_{1\sim{\ }({{N}_{e}}/2)-2}},x{}_{({{N}_{e}}-1)\sim{\ }({{N}_{e}}/2)+2)}\}$$|:(({{y}_{1}},{{x}_{{{N}_{e}}-1}}),$ $\cdots ,({{y}_{({{N}_{e}}/2)-2}},{{x}_{({{N}_{e}}/2)+2}}))$

 (2) In the interval of $[({{{N}_{e}}}/{2}\;)+2,{{N}_{e}}-1]$, (or: $[({{N}_{e}}/2),{{N}_{e}}-1]$, this case is not discussion), when $x,y$ are all positive odd numbers, along the direction of variable x,y delete corresponding even ${{N}_{e}}$ as shown in Figure 4(b), it also has two kind of matching add sun relation.
 %
 %

 $a$) X direction:

 Deleting $\{{{x}_{({{N}_{e}}/2)+2\sim{\ }({{N}_{e}}-1)}},y{}_{({{N}_{e}}/2)-2\sim{\ }1)}\}$$|:(({{x}_{(N/2)+2}},{{y}_{({{N}_{e}}/2)-2}}),$$\cdots ,({{x}_{{{N}_{e}}-1}},{{y}_{1}}))$

 $b$) Y direction:

 Deleting $\{y{}_{({{N}_{e}}/2) \sim{\ }({{N}_{e}}-1)},{{x}_{({{N}_{e}}/2) \sim{\ }1}}\}$$|:(({{y}_{(N/2)}},{{x}_{({{N}_{e}}/2)}}),$$\cdots ,({{y}_{{{N}_{e}}-1}},{{x}_{1}}))$

Only in this way, above even full permutation of $\sum\nolimits_{\varphi }{({{N}_{e}})}$ numbers of matching add sum relationships can be deleted completely. Because when even matrix case: within interval  $[1,{({{N}_{e}}}/{2)-1}]$ and $[({{{N}_{e}}}/{2})+1,{{N}_{e}}-1]$, and when odd matrix case: within interval $[1,{{{N}_{e}}}/{2}\;]$ and $[({{{N}_{e}}}/{2}\;)+2,{{N}_{e}}-1]$, or:$[1,({{{N}_{e}}}/{2}\;)-2]$ and $[({{{N}_{e}}}/{2}\;),{{N}_{e}}-1]$, them two interval range here are certainly prime existence. Obviously, this assumption with the deduction 5.2 and 5.3 are all contradiction.

\textbf{Situation 3.} Assume that in the variables of $x,y$, there are $\pi ({{N}_{e}})\approx {{{N}_{e}}}/{\ln {{N}_{e}}}\;$ prime numbers in normal distribution. If ${{N}_{e}}$ not exist within the matrix $S_{O}^{+}(\max {{N}_{e}})$ = $\max <x,y>$, there is only one possibility that is in the direction of $x,y$, the even adding sum relationships of $S_{O}^{+}(\max {{N}_{e}})$ = $\max <x,y>$ is all can not matching with each other prime numbers, and the combination number as follow.
 \[{{\pi }_{o\leftrightarrow p(or:p\leftrightarrow o)}}({{N}_{e}})=({{N}_{e}}/2)-2({{N}_{e}}/\ln {{N}_{e}})\]

Thus, there are three kinds of cases that can't satisfy the matching results:
\[<{{x}_{i}},{{y}_{j}}>(i,j=2n-1,n\ge 1){{|}_{{{\pi }_{(x,y)}}=({{N}_{e}}/2-2({{N}_{e}}/\ln {{N}_{e}})}}\ne {{N}_{e}}  \tag{5-1}             \]
\[<{{x}_{i}},{{y}_{p}}>(i=2n-1,n\ge 1,{{p}_{{}}}\quad i{{s}_{{}}}\quad prime){{|}_{{{\pi }_{(x,p)}}=({{N}_{e}}/\ln {{N}_{e}})}}\ne {{N}_{e}}    \tag{5-2}         \]
 \[<{{x}_{p}},{{y}_{j}}>(j=2n-1,n\ge 1,{{p}_{{}}}\quad i{{s}_{{}}}\quad prime){{|}_{{{\pi }_{(p,y)}}=({{N}_{e}}/\ln {{N}_{e}})}}\ne {{N}_{e}}   \tag{5-3}          \]

Or even if appear the prime matching, but the number of $\pi _{P(x,y)}^{{}}({{N}_{e}})=({{N}_{e}}/\ln {{N}_{e}})$ can't satisfy the requirement of matching adding sum relationship for the even ${{N}_{e}}$. The result has two different forms that can't meet the matching as follow:
\[<{{x}_{i}},{{y}_{j}}>(i,j=2n-1,n\ge 1){{|}_{{{\pi }_{(x,y)}}\;=\;({{N}_{e}}/2-({{N}_{e}}/\ln {{N}_{e}})}}\ne {{N}_{e}}\]
\[<{{x}_{p}},{{y}_{p}}>({{p}_{{}}}\; i{{s}_{{}}}\; prime){{|}_{{{\pi }_{(p,p)}}\;=\;({{N}_{e}}/\ln {{N}_{e}})}}\ne {{N}_{e}}\]

Because of matching form of the situation 3 makes ${{N}_{e}}$ not exist,  $\sum\nolimits_{\varphi }{({{N}_{e}})}$
even numbers in $S_{O}^{+}(\max {{N}_{e}})$ = $\max <x,y>$  should be all deleted. Obviously, thus result
with the lemma 5.3 is contradiction.

The above three cases can fully prove that if there is an even ${{N}_{e}}$ not exist in the matrix ${{S}_{P}}({{N}_{e}})$, then the result must in $S_{O}^{+}(\max {{N}_{e}})$ all delete $\sum\nolimits_{\varphi }{({{N}_{e}})}$ = ${{N}_{e}}/2$ even numbers, this is impossible, because the matrix $S_{O}^{+}(\max {{N}_{e}})$ does not existence.This is with the lemma 5.1, the lemma 5.2 and other relevant definition conditions are all contradiction. In addition , according to the definition 5.1, 5.2 and the definition 5.3 have known that, because of $S_{O}^{+}(\max {{N}_{e}})$ $\in $ $S_{O}^{+}({{N}_{e}})$ $\in $ $S_{O}^{{}}({{N}_{e}})$, and according to the lemma 5.3 known that, the matrix ${{S}_{P}}({{N}_{e}})$ is the implicit matrix of the matrix $S_{O}^{{}}({{N}_{e}})$. Therefore , as long as there is an even ${{N}_{e}}$ not exist in the matrix ${{S}_{P}}({{N}_{e}})$, it means that $\sum\nolimits_{\varphi }{({{N}_{e}})}={{{N}_{e}}}/{2}\;$ not existing, this result is possible. Obviously, $\sum\nolimits_{\varphi }{({{N}_{e}})}={{{N}_{e}}}/{2}\;$ not existence equivalent to the matrix $S_{O}^{+}(\max {{N}_{e}})$ not existing, and the matrix $S_{O}^{+}({{N}_{e}})$ not existing, and even matrix $S_{O}^{{}}({{N}_{e}})$ is not existing, too. Obviously this is a contradiction.

\textbf{Deduction 5.5.} Any sufficiently large positive even ${{N}_{e}}\ge 6$, it certainly existing a sufficiently large matrix $S_{P}^{+\infty }({{N}_{e}})$, and it is countable infinitely extensible and constructable.

\textbf{Proof.}  According to the definition 5.4, the deduction 5.1, the lemma 5.4 and the theorem 5.1, it's easy to prove this corollary is true.

\textbf{Theorem 5.2 (Even number unique existence theorem). } Any given a positive even number ${{N}_{e}}\ge 6$, it must be unique exist in the matrix $Mod\,\overset{\equiv }{\mathop{X}}\,(p)$.
%

\textbf{Proof.} According to the theorem 5.1, it indicates that if any even ${{N}_{e}}$ has not to exist the matrix ${{S}_{P}}({{N}_{e}})$, it must to delete all $\sum\nolimits_{\varphi }{({{N}_{e}})}$ = ${{N}_{e}}/2$ even numbers in the corresponding matrix $S_{O}^{+}(\max {{N}_{e}})$, but this result is impossible. Because to delete $\sum\nolimits_{\varphi }{({{N}_{e}})}$ = ${{N}_{e}}/2$ even numbers, it is equivalent to delete the set of given even ordered pairs of $\{6,8,10,12,\cdots ,{{N}_{e}}=2n,n\ge 3\}$ all within $(6\le x\le 2n={{N}_{e}})$. Once full deletion could be true, the results can but show that $S_{O}^{+}(\max {{N}_{e}})$ not existing is true, the matrix $S_{O}^{{}}({{N}_{e}})$ is also not existing. In fact, these deleting  with above listed relevant conditions and the theorems are all contradictory. Conversely, since in the $S_{O}^{+}(\max {{N}_{e}})$, all deleting given even $\sum\nolimits_{\varphi }{({{N}_{e}})}$ = ${{N}_{e}}/2$ numbers is impossible. Then as long as there is a given even ${{N}_{e}}$ which does not be deleted in full permutation, it explains that the even ${{N}_{e}}$ which the location of rows and columns with it the parameters of the matching value are not belong to deleted corresponding even. On the contrary, if not deleting is true, there are only two parameters of $(x,y)$ which are all composed of prime numbers, then the even ${{N}_{e}}$ certainly existing in the matrix ${{S}_{P}}({{N}_{e}})$. Therefore, when any given positive even ${{N}_{e}}\ge 6$, according to the deduction 5.1 known that every even in the set of all even number sequence, it certain unique existing the matrix $S_{P}^{+\infty }({{N}_{e}})$ that which is countable infinity extensible and constructable. And because the relation of any even adding sum, those are all can be described by the model $Mod\overset{\equiv }{\mathop{X}}\,(p)$ in the definition 3.2, and the model $Mod\,\overset{\equiv }{\mathop{X}}\,(p)$ of the existence form of describe even numbers is equivalent to the expression of the model $S_{P}^{+\infty }({{N}_{e}})$. Therefore, any given  positive even ${{N}_{e}}\ge 6$, it certainly unique exist in $Mod\,\overset{\equiv }{\mathop{X}}\,(p)$.

\textbf{Theorem 5.3.}  The halting problem of the model $G{{N}_{e}}TM$ is not existing.

\textbf{Proof.}  We have know that, according to the theorem 5.1 and the theorem 5.2, if even ${{N}_{e}}$ is not existing in the model $S_{P}^{+\infty }({{N}_{e}})$ (or $Mod\,\overset{\equiv }{\mathop{X}}\,(p)$), it just right explain corresponding to the status ${{q}_{j}}\notin T$ within the model $G{{N}_{e}}TM$. That is to say, suppose the state ${{q}_{j}}\in F,$ it shows that the machine appear halting question (the controller algorithm can be checked). On the contrary, the above proof result tell us to have known that, if the sequence set of even ${{N}_{e}}$ are all to exist within the model $S_{P}^{+\infty }({{N}_{e}})$ $(or\, Mod\,\overset{\equiv }{\mathop{X}}\,(p))$, it exactly corresponding the status ${{q}_{j}}\in T$ in the model $G{{N}_{e}}TM$. At this time, the machine not halting satisfy operation run status as ${{q}_{0}}\to {{q}_{1}}\to {{q}_{2}}$ $\to \cdots {{q}_{j}}\to \cdots $, it continuously moving down and until infinity. Thus,it is impossible that the model $G{{N}_{e}}TM$ appear the halting problem . In practice, the halting problem in this way is not existence. Conversely, this result also has been proved indirectly that for $\forall N{}_{e}\ge 6$ even number, the even Goldbach conjecture is all established.

\section{Conclusion}

In conclusion, according to the characteristics of even Goldbach conjecture, this paper proposes a new Turing machine computing model $G{{N}_{e}}TM$, which can accurately describe even the proposed results of even Goldbach conjecture. By the existence theorem of the even number adding sum, and the judgment analysis of general probability speculation in the model $\bmod \overset{\equiv }{\mathop{M}}\,({{N}_{e}})$ is given. These proof has been got a result, once the prime in P is determined, by randomly selecting the element ${{q}_{j}}$ in Q, if only computing the element numbers value at $r\approx 0.8\text{33}\sqrt{{{N}_{e}}}$, the result get ${{q}_{j}}$ is a prime in a probability of 50\%, and it can satisfy the matching requirements of even Goldbach number $G({{N}_{e}})$.

Secondly, a new computational model Turing machine $G{{N}_{e}}TM$ is designed in the controller's prime matching rule algorithm, which is recursively solvable by all computers.  For $\forall {{N}_{e}}\ge 6$, the proof obtain as the following result, the matching predicate $p(n)$ of even Goldbach Conjecture under limit existence quantifier is closed independent and decidable. The predicate $p(n)$ is a primary recursion, and the prime matching rule algorithm in the controller is also computer recursively solvable.

In the end, we put the computing problem of infinite existence of even Goldbach conjecture develops into equivalent to the halting question in the model $G{{N}_{e}}TM$ as research. By the matrix model of full arranged intuitive construction of given even ${{N}_{e}}$,we have been equivalence proved  any given even ${{N}_{e}}$ unique exist the matrix model ${{S}_{P}}({{N}_{e}})$ and the model $Mod\,\overset{\equiv }{\mathop{X}}\,(p)$, and the conclusion that not existing halting problem in model $G{{N}_{e}}TM$ is direct given. At the same time, at least have one pair prime matching algorithm can satisfy indirectly given the proposition requirement by even Goldbach conjecture, and this paper proves that the results of infinite conjecture have been established.

Therefore, we can also conclude that the construction of a new computational model can be extended to many similar infinite existence problems in the field of number theory. The infinite existence of Mersenne Primes is studied as an example.

\end{document}